\documentclass[journal]{IEEEtran}
\usepackage[utf8]{inputenc}
\usepackage{amsmath}
\newcommand\norm[1]{\left\lVert#1\right\rVert}
\usepackage{amsfonts}
\usepackage{cite}
\usepackage{dirtytalk}
\usepackage[ruled,vlined,linesnumbered]{algorithm2e}
\usepackage{algorithmic}
\usepackage{xcolor}% enables use of color package
\usepackage{colortbl}
\usepackage{lipsum}
\usepackage{bm}
\usepackage{svg}
\DeclareGraphicsExtensions{.png,.pdf}
\usepackage{caption}
\usepackage{subcaption}

\usepackage{graphicx}
\graphicspath{{figures/}}
\usepackage{graphicx}
\usepackage{tabularx,booktabs}
\newcolumntype{C}{>{\centering\arraybackslash}X}
\usepackage{makecell}

\usepackage{subfiles} % Best loaded last in the preamble
\usepackage{multicol}
\usepackage{blindtext}
\usepackage{kantlipsum}
\usepackage{booktabs,threeparttable}

\begin{document}

    \title{An Online Data-Driven Method to Locate Forced Oscillation Sources from Power Plants Based on Sparse Identification of Nonlinear Dynamics \color{black} (SINDy)\color{black}}
    %PMU measurement based Sparse Identification  of  Nonlinear Dynamics for forced oscillation source identification}
    \date{October 2020}
    \author{Yaojie~Cai,
    Xiaozhe Wang,~\IEEEmembership{Senior Member,~IEEE,} 
    Geza Joos,~\IEEEmembership{Life Fellow,~IEEE,} and
    Innocent Kamwa,~\IEEEmembership{Fellow,~IEEE} %
    \thanks{
    %Manuscript received July 20, 2021; revised October 31, 2021 and March 19 2022; accepted July 3, 2022. Date of publication September 21, 2021; date of current version December 23, 2021. 
    This work was supported by the Fonds de Recherche du Quebec-Nature et technologies under Grant FRQ-NT PR-298827. (Corresponding author: Xiaozhe Wang.)
    
    %Yaojie Cai, Xiaozhe Wang, and Geza Joos are with the Department of Electrical and Computer Engineering, McGill University, Montréal, QC H3A 0G4, Canada (e-mail: yaojie.cai@mail.mcgill.ca; xiaozhe.wang2@mcgill.ca; geza.joos@mcgill.ca).
    %Innocent Kamwa is with the Department of Electrical and Computer Engineering, Laval University, Québec, QC G1V 0A6, Canada (e-mail: innocent.kamwa.1@ulaval.ca).
    }
    }

    % {Shell \MakeLowercase{\textit{et al.}}: Bare Demo of IEEEtran.cls for IEEE Journals}
    
    \maketitle
        
        % The abstract
    \begin{abstract}
            Forced oscillations may jeopardize the secure operation of power systems. To mitigate forced oscillations, locating the sources is critical.  %requires two critical steps, namely, distinguishing forced oscillation from natural oscillation and locating forced oscillation sources. 
            In this paper, leveraging on Sparse Identification of Nonlinear Dynamics (SINDy), an online purely data-driven method to locate the forced oscillation is developed. Validations in all simulated cases (in the WECC 179-bus system) and actual oscillation events (in ISO New England system) in the IEEE Task Force test cases library are carried out,  %Simulation studies %\color{red}in the IEEE 68-bus tbd system, \color{blue} 
            %in the WECC 179-bus system, \color{black} and in the actual oscillation events in ISO New England 
            which demonstrate that the proposed algorithm, requiring no model information, can accurately locate sources in most cases, even under resonance condition %when the forcing strength is low, and 
            and when the natural modes are poorly damped. The little tuning requirement and low computational cost make the proposed method viable for online application. \color{black}
    \end{abstract}
    % keywords 
    
%MAIN%%%%%%%%%%%%%%%%%%%%%%%%%%%%%%%%%%%%%%%%%%%%%%%%%%%%%%%%     
    \section{\uppercase{Introduction}}
        %\color{blue} 12 apr 2021 \color{black}
        
        \IEEEPARstart{S}{ustained} oscillation is one of the major threats to the security and stability of electric power systems, which may introduce undesirable vibrations, damage power grid equipment, and interrupt power supply. In the most severe scenario, growing oscillations may lead to catastrophic cascading blackouts\cite{kosterev1999model}. 
        
        Sustained oscillations can be classified into natural oscillations and forced oscillations. Natural (modal) oscillations may result from high-gain fast exciters, malfunction of control devices, weakly tied transmission lines, and load fluctuations \cite{5604719}, while forced oscillations may appear when power systems are excited by external periodic disturbances due to large cyclic loads, erroneous control loops in power plants, etc.  \cite{rostamkolai1994evaluation}.  Unlike natural oscillations that have been extensively studied in many previous works (e.g.,  \cite{kundur1994power, trudnowski2009overview, vanfretti2011application}), forced oscillations have gained increasing attention recently.
        %\color{blue}
        %as our ability to capture more precise details of the new or old oscillation modes increases with the growing PMU coverage. %is expected to increase \cite{guideline2016pmu}, more precise details of the new or old oscillation mode are expected to be explored.
        % (R1-1) 
        % \color{black}
        IEEE PES Task Force on Oscillation Source Location was established in 2016 as many ``new"  oscillatory behaviors captured in power grids that could not be reproduced through traditional natural oscillation analysis. \color{black} Note that the oscillatory behaviors themselves may not be new but were captured recently owing to an increasing coverage of \color{black} Phasor Measurement Units (PMUs). %(R1 - Introduction 1 )  
        \color{black} 
        Further investigations suggested that they were forced oscillations introduced by specific generators, either under special conditions or with failed components \cite{PEStfIEEE}. These observations clearly show that mitigating forced oscillations are significant, imperative yet challenging to ensure the secure operation of power systems. The most effective means to mitigate forced oscillation is to locate and remove/correct the sources.
        With the help of synchrophasor technology that provides accurate synchronized measurements with high sampling frequencies, %to the control center, providing essential information for real-time monitoring and control of modern power systems. 
        many forced oscillation locating methods
        %detection methods 
        have been developed using PMU data.  %spectrum-based methods have been developed in \cite{wang2015data} \cite{tang2019periodogram}, which can identify forced oscillations from natural oscillations (limit cycles) by exploiting the physical fundamental difference between the two oscillation mechanisms. 
        \color{black}
        %\color{black} 
        %In spite of big advancement, the method requires accurate knowledge of transmission network, which may not always be available in practice. %makes strong assumptions like a lossless network and the constant power load model. %\color{red} constant impedance or constant power load model?\color{black}  
        Transfer function based methods have been proposed in \cite{ 8273770}, which identify sources of forced oscillations from indexes derived from the assumed system transfer function. Effective impedance based methods were developed in \cite{Impedance} to locate forced oscillation sources by comparing the measured current spectrum of system components with predicted ones. However, %both the transfer function based methods \cite{7935743, 8273770} and the effective impedance based methods \cite{Impedance, Bayesian} 
        all the aforementioned works %\color{red} can you please confirm? \color{blue} (\cite{7935743, Bayesian} might need knowledge of generator model to construct frequency response function (FRF) or transfer function)   \color{black} 
        require accurate information of power network %or \color{red} is it ``and" or ``or"?\color{black} \color{blue} (11June2021) I think it is "and". Because the generator model is important for both algorithm, both of them is relay on Frequency Response Function (FRF) which is derived from the generator model. 
        %(FRF) \color{black} 
        and generator model parameters, which may not always be available in practice due to communication errors, bad data, frequent line switching, etc. To relax the assumption of accurate knowledge of power network and dynamic model,  purely data-driven methods for forced oscillation locating have been recently developed in \cite{xu2020location, foHuang}. The concept of oscillation phasor estimated from PMU data is utilized in \cite{xu2020location} to locate the forced oscillation source, yet the performance of the method under resonance or under low forcing strength is in question as the estimation of oscillation phasor and damping ratio may be deteriorated. %\color{blue} (15dec2020) it is heavily depend on the damping ratio estimation from the osillation phasor, and the threshold between forced oscillation and weak damped natural oscillation is set as 2 percent. So any weak damped natural oscillation with ratio is lower than 2 percent is deemed as forced oscillation. Judge by the area of half cycle in ROCOF plot waveform, the frequency changes is between 0.01 Hz which is much bigger than our case.    \color{red} confirm if it is correct\color{black}. 
        The authors of \cite{foHuang} %proposed a compressed sensing approach 
        applied Robust Principal Component Analysis (RPCA) to extract the sparse and low-rank components of the measurement matrix, which are linked to the resonance-free and resonance components of system dynamics, %yet the mapping lacks without theoretical proof. 
        while the resonance-free component is used to locate forced oscillations. Despite being intuitive, the theoretical proof of the mapping between the physic model and the low-rank/sparse components remains to be seen. The authors of \cite{EnergyBased} proposed the Transient Energy Flow (TEF) method that identifies the oscillation source as the one that generates oscillation energy, which has been well tested in \cite{ISONewEnglandExperience} with great success.
        \color{black}
        However, as discussed in \cite{foHuang}, the TEF method may fail if the source generator is not monitored and erroneous system topology is used. \color{black} %\color{red} any other comments about RPCA? I think you mention that for some cases (FM), it may not work well? Why?\color{blue} (add remarks in the validation section) 
        In this paper, we propose a purely data-driven method to locate forced oscillation sources. %\color{green} (delete) in near real time \color{black}. %forced oscillation detection and locating method is proposed. 
        Leveraging on the methodology of Sparse Identification of Nonlinear Dynamics (SINDy) \color{black} \color{black}\cite{brunton2016discovering, zhang2019convergence, brunton2016sparse, kaiser2018sparse} \color{black} that combines compressed sensing technique with nonlinear dynamical systems, we show that the power system dynamic model can be extracted from PMU measurements. As the extracted dynamic equations have a clear physical interpretation, the location of forced oscillation sources can be carried out using the extracted model. The performance of the proposed method has been tested in all 24 cases available in the IEEE Task Force test cases library \cite{realCasePaper, websiteForcedOsi} (18 simulated cases in the WECC 179 bus system and 6 actual oscillation events in ISO New England (ISO-NE) system) %\cite{realCasePaper, websiteForcedOsi} 
        using raw PMU data. %\color{green} (delete) without pre-processing \color{black}. %Particularly, compared to RPCA method in \cite{foHuang}, the proposed method seems to be more accurate in the cases where the forced signals come from turbine governors. 
        %\color{red} comparison with RPCA. also is very effective if the source of oscillation is from turbine governor. 
        %\color{blue} Compare to other low order approximation method, the proposed algorithm is very effective if the source of oscillation is from turbine governor. \color{black} 
        In sum, the contributions of this paper are presented as below:
        \begin{enumerate}
            \item The proposed method is purely data-driven, requiring no information of the dynamic model and the network parameters. 
        %    \item The proposed method can accurately distinguish the forced oscillation from natural oscillation, while locating the forced oscillation sources, even under resonance condition, and/or when the forcing strength is low, and/or when the natural modes are poorly damped.
            \item The proposed method, requiring a little tuning of parameters, and trivial computational cost, possesses good feasibility in practical online applications.
            \item The proposed method can accurately locate forced oscillation sources in most of the cases available in the library including resonance cases, low forcing strength cases, rectangular forcing signal cases, etc. 
            \item Compared to the RPCA method in \cite{foHuang}, the proposed method seems to be more accurate in the cases where the forced signals come from turbine governors of generators, serving as a good complement to the RPCA method. \color{black} 
            \color{black} Compared to the Dissipating Energy Flow method in \cite{ISONewEnglandExperience}, the proposed method may help pinpoint the generators closest to the source generator when the source generators are not monitored, providing useful guidance for source searching.  
        %    \item If the source generator is not monitored, the proposed method pinpoint the closet generator as the result which provides a good guide for source finding. (R1  <Introduction> 2) \color{black}
            %\item The potential reasons for the failure of the method in two cases are discussed.  The proposed algorithm requires little tuning of parameters and can use raw PMU data without pre-processing (e.g., detrending, filtering). 
        \end{enumerate}
        In addition, the potential reasons for the failure of the method in two cases are discussed. 
        The test results imply that the performance of the proposed method in cases where the forced oscillations are \color{black}completely unobservable \color{black} from measurements \color{black} due to, for example, extremely low forcing strength %which is due to the close distance between sensor unit (R1 <Introduction> 4.) 
        \color{black} may deteriorate. 
        Besides, since the library considers only the cases where the source of forced oscillations are from generators (i.e., exciters, turbine governors), future efforts are needed to extend and test the proposed method when the forced oscillation sources are not generators (e.g., HVDC control). 
        \color{black}
        \color{black}
        
%%%%%%%%%%%%%%%%%%%%%%%%%%%%%%%%%%%%%%%%%%%%%%%%%%%%%%%%%%%% 
    \section{\uppercase{power system dynamic model for forced oscillation analysis}} \label{section:dynamic model}

        The power system dynamics can be described as a set of nonlinear Differential Algebraic Equations (DAE). 
        Particularly, power generator rotor dynamics \color{black}play critical roles \color{black} in ambient conditions \color{black} \cite{PalCoherency, wang2017pmu,shengOnlineDyn}\color{black}, %\color{red} cite power system coherency and model reduction, Analysis of Power System Oscillations for
        %Developing Synchrophasor Data Applications,  PMU-Based Estimation of Dynamic State Jacobian
        %Matrix and Dynamic System State Matrix in Ambient Conditions\color{black}, 
        which can be represented by the classical swing equations: 
        %Particularly, the power generator rotor dynamics are \color{blue}one of the critical elements to the dynamic in ambient conditions, which the classical swing equations can represent: \color{black}
        
        \vspace{-3ex}
        \normalsize 
        \small
        \begin{equation} \label{eq:swingEquation_1}
                \begin{split}
                    \bm{\dot{\delta}} &= \bm{\omega} \\
                    M\bm{\dot{\omega}} &=  \bm{P_{m}} - \bm{P_{e}} - {D} \bm{\omega}+\bm{u}
                \end{split}
        \end{equation}
        \normalsize 
        where $ \bm{\delta} = [\delta_1, \cdots, \delta_r]^T$; 
        $\bm{\omega} = [ \omega_1, \cdots, \omega_r]^T$ are the vectors of rotor angles and angular frequencies of all generators; 
        $M=\mbox{diag}[M_1, \cdots, M_r]$ is a diagonal matrix containing the inertia constants of all generators; $D=\mbox{diag}[D_1, \cdots, D_r]$ contains the damping coefficients of all generators; $\bm{P_{m}} = [P_{m_{1}}, \cdots,  P_{m_{r}}]^T$ and $\bm{P_{e}} = [P_{e_{1}}, \cdots, P_{e_{r}}]^T$ denote the vectors of mechanical power and electrical power of generators; $\bm{u}$ denotes the external input. In forced oscillation study, $\bm{u}$ may represent the periodic forcing imposed on generator shaft due to, for example, inappropriate parameters in the control loop from turbine governor or exciter. \color{black} Indeed, as shown in Appendix \ref{app:rotorDynamics}, the forced oscillations from either excitation systems or turbine governors may manifest themselves in swing equations given in (\ref{eq:swingEquation_1}). \color{black} % Although exciter dynamics are indirectly coupled with frequency and rotor angle through field flux \cite{kundur1994power}, the forced oscillation from exciters may still be manifested in rotor dynamics, which will also be shown in Section \ref{section:numerical study}. \color{black}
        Due to its periodic nature, \color{black}
        $\bm{u}(t)=[u_1(t),...u_{r}(t)]^T$
        \color{black} can be described by Fourier series. \color{black} For each $u_j(t)$, $j\in\{1,2,...,r\}$, it can be represented as:
        %\color{red} use ``eqnarray" to make $=$ aligned. would it be better to use $\zeta$ instead of $\sqrt{\zeta}?$ 
        
        \vspace{-4ex}
        \color{black}
        \small
            \begin{equation} \label{eq:inputFO}
               \begin{split}
                    %\bm{u}\left(t \right)=\sum_{i=1}^{l}\Bigl( a_i \sin \left(\omega_{F_{i}} t\right)+ 
                    %b_i \cos \left(\omega_{F_{i}} t\right)\Bigr) \\
                    {u}_j\left(t \right) & = \sum_{i=1}^{l}\Bigl( a_{ij} \sin \left(\omega_{F_{ij}} t\right)+ 
                    b_{ij} \cos \left(\omega_{F_{ij}} t\right)\Bigr)\\ 
                    &=\sum_{i=1}^{l}\Bigl(\sqrt{a_{ij}^2+b_{ij}^2}\sin(\omega_{F_{ij}}t+\varphi_{ij}\Bigr)\\
                 %   & = \sum_{i=1}^{l}\Bigl( \underbrace{\sqrt{\zeta_i} \cos (\varphi)}_{=a_i} \sin (\omega_{F_{i}} t)+\underbrace{\sqrt{\zeta_i} \sin (\varphi)}_{= b_i } \cos (\omega_{F_{i}} t) \Bigr) \\
                    & = \sum_{i=1}^{l}\Bigl( \sqrt{\zeta_{ij}} \sin \left(\omega_{F_{ij}} t + \varphi_{ij} \right)\Bigr)   
                \end{split} 
            \end{equation}
        \normalsize 
        where $\zeta_{ij}={a_{ij}^2+b_{ij}^2}$, $\varphi_{ij}=\arctan(\frac{b_{ij}}{a_{ij}})$. \color{black}
        Besides, the electrical power injection to the grid from generator $i$: 
        
        \vspace{-2ex}
        \small
            \begin{equation}\label{eq:pe}
                P_{e_{i}} = E^{2}_{i} G_{ii} + \sum_{i=1, j \neq i}^{r} E_{i}E_{j} Y_{ij} \cos{ \left(\phi_{ij} - \delta_{i} + \delta_{j} \right)} \vspace{-2ex}
            \end{equation} 
        \normalsize 
        where $E_{i}$ is the $i$th generator internal electromotive force (emf) magnitude. 
        $Y_{ij}\angle \phi_{ij}=G_{ij}+jB_{ij}$ corresponds to the $(i,j)^{th}$ entry of the reduced admittance matrix. 
        As discussed in \cite{odun2013analysis}, 
        power systems are naturally experiencing random perturbations from load power variations, which can be reflected in the diagonal
        elements of the equivalent admittance matrix $Y$ seen from the generator internal buses:
        
        \vspace{-3ex}
        \small
                \begin{equation} \label{eq:load}
                    \begin{split}
                    Y\left(i, i\right) = Y_{ii} \left( 1 + \sigma_{load_{i}}{\eta_{i}} \right) \angle \phi_{ii} \vspace{-5ex}
                    \end{split} 
                \end{equation}
        \normalsize
        where $\sigma_{load_{i}}$ is the standard deviation of load variation. $\eta_{i}$ is a standard Gaussian random variable. 
        \color{black}
        %Even-through the small load power factor changes cause $\phi_{i i}$ varies; however, it has a limited impact on the generator rotor dynamics.
        It is typically assumed that the random variation does not affect $\phi_{i i}$  such that load power factor remains constant  \cite{odun2013analysis} \color{black} in a time window of minutes considering the fact that power factor correction is carried out in hourly time scale.  \color{black}
        Thus, the swing equation considering stochastic load variation can be represented as \cite{wang2017pmu} \cite{9115088}: 
        
        \vspace{-2ex}
        \small
                \begin{equation} \label{eq:swingEquation_org}
                    \begin{split}
                        \bm{\dot{\delta}} &= \bm{\omega} \\
                        M\bm{\dot{\omega}} &=   \bm{P_{m}} - \bm{P_{e}} - {D} \bm{\omega} - {E}^{2}{G} \Sigma \bm{\eta} +\bm{u} \vspace{-2ex}
                    \end{split}
                \end{equation}
        \normalsize
        where $E=\mbox{diag}[E_1, \cdots,  E_r]$; ${G}=\mbox{diag}[G_{11}, \cdots,  G_{rr}]$; 
        ${\Sigma}=\mbox{diag}[\sigma_{load_{1}},  \cdots,   \sigma_{load_{r}} ]$; $\bm{\eta}=[\eta_1,  \cdots,  \eta_r ]^T$; 
        \color{black} 
        \color{violet}
        %The assumption that rotor dynamics are dominant dynamics in ambient conditions has been widely used in system-wide electromechanical oscillation studies such as inter-area oscillation and coherency identification.\cite{pierre1997initial, dosiek2012mode, PalCoherency, chakrabortty2012wide }.
        %As discussed in \cite{vanfretti2010, chow2013power},
        \color{black}
        Similar to \cite{vanfretti2010, Impedance}, the classical generator model is considered in this study as we are interested in system dynamics in ambient conditions.  \color{black}%generator rotor dynamics are the dominant dynamics in ambient conditions. \color{black} %\color{blue} one of the critical roles in ambient conditions dynamics. \color{black} %as discussed in  \color{violet} \cite{Impedance} \color{red} cite  L. Vanfretti and J. Chow, “Analysis of power system oscillations for
        %developing synchrophasor data applications,” in Proc. IREP Symp. Bulk
        %Power Syst. Dynamics and Control VIII, Rio de Janeiro, Brazil, 2010. \color{blue}. However, 
        %Although the classical generator model is considered when the algorithm is developed, 
        Nevertheless, the performance of the proposed method is tested when higher-order generator models with detailed control devices are implemented, as well as in the actual oscillation events with no information about the dynamical model, as shown in Section \ref{section:numerical study}-\ref{section:real case testing}. Besides, the performance of the method is also tested under load power factor variations as presented in Section  \ref{section:powerFactorChange}. \color{black}These results further confirm that the rotor dynamics described in (\ref{eq:swingEquation_org}) are sufficient for locating forced oscillation sources. \color{black}
        
        Under small perturbation, (\ref{eq:swingEquation_org}) can be linearized around its steady-state operating point as below: 
        
        \small
            \begin{equation} \label{eq:swingEqVect}
            \begin{split}
                \left[\begin{array}{c}
                \Delta\bm{ \dot{\delta}} \\
                \Delta\bm{ \dot{\omega}}
                \end{array}\right]=\left[\begin{array}{cc}
                0_{r\times r} & I_{r\times r} \\
                -M^{-1} \frac{\partial \bm{P_{e}}}{\partial \bm{\delta}} &-M^{-1} D
                \end{array}\right]
                \left[\begin{array}{c}
                \Delta\bm{ \delta} \\
                \Delta\bm{ \omega}
                \end{array}\right] \\
                +\left[\begin{array}{c}
                \bm{0_{r\times 1}} \\
                {-M^{-1}} {E}^{2} {G} \Sigma \bm{\eta}
                \end{array}\right] + 
                \left[\begin{array}{c}
                \bm{0_{r\times 1}} \\\bm{u}
                 \end{array}\right]
            \end{split}
            \end{equation}
        \normalsize
        As can be seen that the governing dynamic equations consist of linear terms, noise terms, and forced oscillation terms, i.e., the three terms on the right-hand side of (\ref{eq:swingEqVect}), respectively. 
        %\color{blue} More discussions about representing the forced oscillation in rotor dynamics are addressed in Appendix \ref{app:rotorDynamics}.\color{black}
        
        \color{blue}
        \color{black}
        
        In this paper, we will exploit the structure of power system physical model to develop a novel method based on SINDy %and the structure of power system physical model 
        to estimate the terms  particularly corresponding to the forced oscillations %\color{blue} ($\bm{a}_{i}$, $\bm{b}_{i}$) \color{black} 
        purely from PMU measurements.
        The magnitude of the terms can be used to locate the sources of forced oscillations. 
        In the next section, the essence and theoretical basis of SINDy method will be briefly introduced.

%%%%%%%%%%%%%%%%%%%%%%%%%%%%%%%%%%%%%%%%%%%%%%%%%%%%%%%%%%%%    
    \section{\uppercase{Sparse Identification of Nonlinear Dynamics}}
        %(\ref{eq:swingEqVect}) shows that the behaviour of the power system can be described by a system consists with a set of first order nonlinear ordinary differential equations. If the derivatives of the states variables in the system are not function of time, then such system is consider as the autonomous with a vector-matrix notation as following: 
        
        %\color{red} where is the general dynamical system? You need to introduce the generic dynamical system you are working on. Make the description of the model complete.  \color{blue} - added the descriptions of the system\color{black}
        Let's consider a generic dynamical system of the following form: 
        \small
        \begin{equation}
            \dot{\bm{x}}=\bm{h}(\bm{x})
        \end{equation}
        \normalsize
        where $\bm{x}$ is the vector of state variables; $\bm{h}$ are the governing equations describing the dynamics of the system. %are stored in the column vector $x$. 
        %\color{red} please explain that why the SINDy algorithm works. some explanation about the combination of symbolic regression and sparse representation. \color{black} 
        The core observation is that for many dynamical systems including power systems, the function $\bm{h}$ is sparse in the space of possible functions (\color{black} more explanation after (\ref{eq:systemSINDy}) for power system model\color{black}). %i.e., consisting of only a few terms.
        %\color{blue}
        %For example, an aerodynamics model for a prototype bridge has a few terms in the function space of linear and high order polynomial \cite{li2019discovering}.
        %\color{black}
        %For example, the power system dynamic equations with forced oscillations may have only a few terms in the space of linear functions and trigonometric functions (see %(\ref{eq:swingEqVect}), (\ref{eq:inputFO}), and \color{blue}
        %(\ref{eq:systemSINDy}). \color{black})
        %(see(\ref{eq:systemSINDy}) (R5-2) ). 
        %Recent advancements in compressive sensing and sparse regression ensure that the sparse solution is found with high probability using convex methods, which can be applied in large-scale systems\cite{brunton2016discovering}. The resulting sparse model identification can nicely balance the model complexity and accuracy. If we consider data measurements 
        In light of this, a sparse identification method of finding the undergoing dynamics SINDy is introduced in \cite{brunton2016discovering}, which can identify the governing equations from measurement data by leveraging advances in sparsity techniques and machine learning. The exploited sparse formulation nicely balances the model complexity (i.e., the sparsity of governing equations) and accuracy, avoiding overfitting the model to the data.  
        
        In particular, given measurement matrix ${X}^T=[\bm{x}(t_1),...,\bm{x}(t_m)]$ and its time derivative matrix $\dot{{X}}$ either measured or numerically approximated, %such as using simple finite difference approximation or other complex method \cite{chartrand2011numerical},  
        $\dot{{X}}$ can be represented as a linear combination of columns from a feature library $\Theta$:
        
        \small
        \begin{equation} \label{SINDyFormat}
            \dot{X}=\Theta({X})\Xi
        \end{equation}
        \normalsize
        where  the feature library
        
        \small
        \begin{equation} \label{general library}
                    \Theta({X})= 
                    \left[
                    \begin{array}{ccc}
                    |  & \dots & | \\
                    \Theta_{1}(X) & \ldots & \Theta_{l}(X) \\
                    | & \dots & | 
                    \end{array}\right]
         \end{equation}
        \normalsize 
        may consist of candidate linear or nonlinear functions such as constants, polynomials, trigonometric functions, etc. $\Xi=\bm{[\xi_1, \xi_2,...,\xi_v]}$ contains the sparse vectors of coefficients to be determined. Once ${\Xi}$ is determined, the dynamics of each state variable can be represented as \color{black} (\ref{eq:sindy1}). \color{black}
        
        \small
        \begin{equation}
            \bm{\dot{x}_k=\bm{h_k(x)}}=\Theta(\bm{x}^T)\bm{\xi_k} \qquad k=\{1,2,...,v\}
            \label{eq:sindy1}
        \end{equation}
        \normalsize
        Note that ${\Theta(\bm{x}^T)}$ is a vector of symbolic functions of $\bm{x}$ while ${\Theta({X})}$ is a matrix.
        Solving $\bm{\xi_k}$ requires distinct optimization for each $k$. More details will be discussed in Section \ref{section:proposed algorithm}. \color{black}SINDy framework has been further extended to include external inputs and feedback (e.g., \cite{brunton2016sparse, kaiser2018sparse}) %control\cite{brunton2016sparse} and combine with model predictive control \cite{kaiser2018sparse}.
        \color{black} %control strategy  \cite{brunton2016sparse} \cite{kaiser2018sparse}, or work with other optimization methods for finding the potential non-linear model between observational data and a hyper-rich and extensive library such as  conditional gradient descent algorithm  \cite{carderera2021cindy}. (16Dec2021)
        The choice of basis functions $\Theta$ is crucial to achieve the sparse dynamics, which nevertheless may not be clear for general dynamical systems and may even require advanced algorithms in machine learning to extract features \cite{brunton2016discovering}. However, in the forced oscillation study of power systems, a reasonable choice of nonlinear functions can be selected because of the periodic nature of forced oscillations and the physical model of power system as discussed in Section \ref{section:proposed algorithm}. %the physical model of power systems provide valuable insights regarding a reasonable choice of nonlinear functions as discussed in the next Section.  %the physical models provide important insights choice of basis functions can be  

        In short, SINDy method includes building a library of basis functions for seeking a linear relationship between the basis functions and the time derivatives of measurement data, %the measurement data and corresponding time derivatives
        and a sparse regression method to estimate the coefficient matrix $\Xi$ that balances sparsity and accuracy. %be discussed in the next section. %which use a sequential thresholded least square fitting for finding a numerical estimation of the relationship matrix with a great sparsity. 
        %\color{red} include more detailed theoretical results about the algorithm \color{black}
        %\color{blue}
        %SINDy framework can be further extended with control strategy \cite{brunton2016sparse} \cite{kaiser2018sparse}, or work with other optimization methods for finding the potential non-linear model between observational data and a hyper-rich and extensive library such as  conditional gradient descent algorithm  \cite{carderera2021cindy}. %(16Dec2021)
        %\color{black}

%%%%%%%%%%%%%%%%%%%%%%%%%%%%%%%%%%%%%%%%%%%%%%%%%%%%%%%%%%%%      
    \section{\uppercase{A data-driven method to locate forced oscillations}}\label{section:proposed algorithm}
    %{SINDy for power system forced oscillation identification}}
        \color{black}
        
        If we substitute $\bm{u}(t)=[u_1(t),...u_{r}(t)]^T$ in (\ref{eq:inputFO}) to the linearized power system stochastic dynamic model,  (\ref{eq:swingEquation_org}), %where 
        %\small
        %    \begin{equation} 
        %    \label{eq:uj}
        %    {u}_j(t) =\sum_{i=1}^{n}\Bigl( a_{ij} \sin \left(\omega_{F_{i}} t\right)+ 
        %            b_{ij} \cos \left(\omega_{F_{i}} t\right)\Bigr)
        %\end{equation}
                    %$j\in\{1,2,...,r\}$, %where , %given in (\ref{eq:inputFO}) 
                    %to the linearized power system stochastic dynamic model (\ref{eq:swingEqVect}
        %\normalsize
                    then the power system model can be represented as:
        %\color{blue}
        
        \vspace{-2ex}
        \small
            \begin{equation} \label{eq:swingEqVec2}
            \begin{split}
                \left[\begin{array}{c}
                \Delta\bm{ \dot{\delta}} \\
                \Delta\bm{ \dot{\omega}}
                \end{array}\right]
                =
                \left[\begin{array}{cc}
                0_{r\times r} & I_{r\times r} \\
                -M^{-1} \frac{\partial \bm{P_{e}}}{\partial \bm{\delta}} &-M^{-1} D
                \end{array}\right]
                \left[\begin{array}{c}
                \Delta\bm{ \delta} \\
                \Delta\bm{ \omega}
                \end{array}\right] \\
                + 
                 \left[\begin{array}{c}
                \bm{0_{r\times 1}} \\
                {-M^{-1}} {E}^{2} {G} \Sigma \bm{\eta}
                \end{array}\right]
                + 
                \left[\begin{array}{c}
                \bm{0} \\ \bm{a}_{1}
                 \end{array}\right] 
                \sin \left(\omega_{F_{1}} t\right) 
                  + \\
                \left[\begin{array}{c}
                \bm{0} \\ \bm{b}_{1}
                 \end{array}\right] 
                \cos \left(\omega_{F_{1}} t\right) 
                 + \dots +
                \left[\begin{array}{c}
                \bm{0} \\ \bm{b}_{n}
                 \end{array}\right] 
                \cos \left(\omega_{F_{n}} t\right) 
            \end{split}
            \end{equation}
        \normalsize
        where $\omega_{Fi}$, $i\in\{1,2,...,n\}$ are the dominant forced oscillation frequencies. It can also be represented \color{black} in  \color{black} %\color{blue} (If you think delete is okay, I will redraw the flow chart) \color{black}
        a compact form:
        
        \vspace{-2ex}
        \normalsize   
        \small
                \begin{equation}
                \begin{split}
                    %\dot{X}^{T} =
                \left[\begin{array}{c}\Delta \dot{\bm{\delta}}\\\Delta \dot{\bm{\omega}}\end{array}\right]=
                    \left[ \begin{array}{cccccc}
                        \bm{0}  &  -M^{-1} E^{2} G \Sigma \bm{\eta} \\
                        0 & -M^{-1} \frac{\partial \bm{P_{e}}}{\partial \bm{\delta} }\\
                        I & -M^{-1}D \\
                        \bm{0} & \bm{a}_{1}\\
                        \bm{0} & \bm{b}_{1}\\
                        \vdots & \vdots\\
                        \bm{0} & \bm{a}_{n}\\
                        \bm{0} & \bm{b}_{n}
                    \end{array}\right]^{T}
                    \left[\begin{array}{c}
                        1 \\
                        \Delta \bm{\delta} \\
                        \Delta \bm{\omega} \\
                        \sin \left(\omega_{F_{1}} t\right) \\
                        \cos \left(\omega_{F_{1}} t\right) \\
                       \vdots \\
                        \sin \left(\omega_{F_{n}} t\right) \\
                        \cos \left(\omega_{F_{n}} t\right)
                    \end{array}\right]
                   \end{split}
                   \label{eq:systemSINDy}
                \end{equation}
        \normalsize
        %\color{red} Can you change the order of the terms for $1$, $\Delta \delta$, $\Delta \omega$? 
        where $\bm{a}_i=[a_{i1},...,a_{ir}]$, $\bm{b}_i=[b_{i1},...,b_{ir}]$, %\color{violet} ( It should be $r$ as the total generator number because I was trying to say for $i$th frequency, there is $1$ to $r$th magnitude, which corresponds to each generator.) (16 Dec 2021)\color{red} I think $a_{ir}$ should be $a_{in}$, same for $b_{ir}$. please check \color{blue} 
        describing the magnitudes of potential forced oscillation inputs. \color{black}This formulation indicates that the power system dynamic model $\bm{h}$ is sparse in the function space of zero-degree polynomial bias, linear functions and trigonometric functions. 
        %, $j\in\{1,2,...,2r\}$. 
        %(R5-1) 
        %\color{blue} Because the matrix 
        %\color{red}TBD\color{blue}In addition since one of the assumptions we proposed is that after the initial forced oscillation injection period, the power system enters a new steady-state with forced oscillation as an input.
        %In case the generator rotor angle and speed are not time-synchronously measured in some cases.
        %The measurements of the terminal voltage angles and frequencies with same common reference time origin as the PMU data are valid replacements to the rotor signal as the application in \cite{modeShapeEst}, and the study of real-life events in Section VI.
        
        %\color{blue}
        %In some cases, the generator rotor angle and speed are not time-synchronously measured.
        %The measurements of the terminal voltage angles and frequencies with the same common reference time origin as the PMU data are valid replacements to the rotor signal as the application in \cite{modeShapeEst}, and the study of real-life events in Section VI.
        %\color{black}
        %In addition, the rotor signal can be estimated around steady-state from PMU measurements \cite{zhou2011calibration}, or directly measured \cite{measureRotor}. (23Dec2021)
        \color{black}

        %We assume that the measurements of machine rotor angles can be obtained, for example, from mechanical speed measurement units of synchrophasor  \cite{SELRotorAngle},   
        %thus the measurement matrix and its time derivative are formed as follows: 
        %\small
        We can further define the measurement matrix $X$ as below, where machine rotor angles can be obtained, \color{black}for example, \color{black} from mechanical speed measurement units of synchrophasor.  \cite{SELRotorAngle}, 
        \begin{equation} \label{eq:measurementMatrix}
                \begin{split}
                X =\left[\begin{array}{ccc}
                \Delta \delta_{1}\left(t_{1}\right) & \ldots & \Delta \delta_{r}\left(t_{1}\right) \\
                \vdots & \ddots & \vdots \\
                \Delta \delta_{1}\left(t_{m}\right) & \ldots & \Delta \delta_{r}\left(t_{m}\right)
                \end{array}\right.
                \end{split}
            \end{equation}    
                $\qquad  \qquad  \qquad \left.
                \begin{array}{ccc}
                \Delta \omega_{1}\left(t_{1}\right) & \ldots & \Delta \omega_{r}\left(t_{1}\right) \\ 
                \vdots & \ddots & \vdots \\
                \Delta \omega_{1}\left(t_{m}\right) & \ldots & \Delta \omega_{r}\left(t_{m}\right)
                \end{array} \right]$
            \begin{equation}
                \begin{split}
                \dot{X} =\left[\begin{array}{ccc}
                \Delta \dot{\delta}_1\left(t_{1}\right) & \ldots & \Delta \dot{\omega}_{r}\left(t_{1}\right) \\
                \vdots & \ddots & \vdots \\
                \Delta \dot{\delta}_1\left(t_{m}\right) & \ldots & \Delta \dot{\omega}_{r}\left(t_{m}\right)
                \end{array}\right]
                \end{split}
            \end{equation}
        \color{black}
        In case the generator rotor angles and speeds are not time-synchronously measured, the measurements of the voltage angles and frequencies from PMU data are good replacements to the rotor signals as discussed in \cite{modeShapeEst}. %and the study of real-life events in Section \ref{section:real case testing}. (R2-7) (R5 - 3) (16Dec2021)\color{red} not clear 
        %\color{black}
        %\color{blue}
        %As discussed in \cite{modeShapeEst}, the measurements of the terminal voltage angles and frequencies from PMU data are valid replacements to the rotor signals.
        \color{black} This is also validated by the results of actual oscillation events presented in Section VI, in which voltage angles at generator terminal buses from PMUs are directly applied. Besides, %as discussed in \cite{zhou2011calibration},  
        %In case the 
         $\bm{\delta}$ and $\bm{\omega}$ in ambient conditions %are not time-synchronously measured, they 
        may also be estimated from PMU measurements \cite{zhou2011calibration}, or directly
        measured \cite{measureRotor}.  \color{black}
        %Besides, 
        %In addition since one of the assumptions we proposed is that after the initial forced oscillation injection period, the power system enters a new steady-state with forced oscillation as an input.
        %In case the generator rotor angle and speed are not time-synchronously measured in some cases.
        Lastly, if $\Delta\dot{\bm{\omega}}$ is unavailable, a finite difference approximation %\color{green} (delete) or other advanced methods (e.g. Total Variation Regularized Numerical 
        %Differentiation (TVDiff) \cite{chartrand2011numerical}) \color{blue} (reason: I tried, the result is really bad) \color{black}
        can be applied to $\Delta\bm{ \omega}$ for estimating $\Delta\dot{\bm{\omega}}$.
        
        Then using the measurement matrix $X$ and taking   transpose  for  both  sides   of  (\ref{eq:systemSINDy}), we can represent the power system model in the form of %a formulation (\ref{SINDyFormat}) used in SINDy:
        %\small
        %\begin{equation} %\label{SINDyFormat}
            $\dot{X}=\Theta({X})\Xi$, i.e., the formulation used in SINDy, 
        %\end{equation}
        %\normalsize
        where the feature library %for SINDy method 
        can be selected naturally as:
        
        \vspace{-3ex}
        %Thus, we select the following feature library for SINDy method: 
        \small
            \begin{equation} \label{library}
                \begin{split}
                    \Theta(X)= 
                    \left[
                    \begin{array}{cccccc}
                    | & | & \dots & | & | & \dots \\
                    \bm{1}  &  \bm{\Delta \delta_{1}} & \dots & \bm{\Delta \delta_{r}} & \bm{\Delta \omega_{1}} & \dots  \\
                    | & | & \dots & | & | & \dots  \\
                    \end{array}\right.
                \end{split}
            \end{equation}
            \vspace{1ex}
            $ \left.
            \begin{array}{ccccc}
            | & \dots & | & | & \dots \\
            \bm{\Delta \omega_{r}} & \dots & 
            \bm{\sin (\omega_{F_{i }}t)}  & 
            \bm{\cos (\omega_{F_{i }}t)}  & 
            \dots \\
            | & \dots &| & | & \dots
            \end{array}\right]_{m \times (1+2r+2n)}$ \\ 
        \normalsize
        %    such that the power system model can be represented as $\dot{X}=\Theta({X})\Xi$.
        %
        and, in theory, the coefficient $\Xi$ matrix should be as follows: 
        
        \vspace{-3ex}
        \small
            \begin{equation}
                \begin{split}
                \label{eq:Xi}
                \Xi =\left[\begin{array}{lll}
                \bm{\Xi_{\eta}} & \Xi_{\text {Jacobian}} & \Xi_{a b}
                \end{array}\right]_{2r \times (1+2r+2n) }\\
                \end{split}
            \end{equation}
            \begin{equation} \label{Xi_DCbias}
                \begin{split}
                \bm{\Xi_{\eta}}=\left[\begin{array}{c}
                %| \\
                0 \\
                %| \\
                -M^{-1} E^{2} G \Sigma \eta \\
                \end{array}\right]_{2r \times 1} 
                \end{split}
            \end{equation}
            \begin{equation} \label{Xi_Jacobian}
                \begin{split}
                \Xi_{\text {Jacobian}}=\left[\begin{array}{cc}
                0 & I \\
                -M^{-1} \frac{\partial P_{e}}{\partial \delta} & -M^{-1} D
                \end{array}\right]_{2r \times 2r} 
                \end{split}
            \end{equation}
            \begin{equation} \label{Xi_ab}
                \begin{split}
                \Xi_{a b}^{T} =\left[\begin{array}{ccccccc}
                0 & 
                \cdots & 
                a_{11} &
                \cdots &
                a_{1r}  \\
                \vdots & 
                \cdots &
                b_{11} &
                \cdots &
                b_{1r}  \\
                \vdots  & \ddots & \vdots & \ddots & \vdots \\
                \vdots & 
                \cdots & 
                a_{n1} & 
                \cdots &
                a_{nr} \\
                0 & 
                \cdots & 
                b_{n1} &
                \cdots &
                b_{nr} 
                \end{array}\right]_{2n \times 2r} 
                \end{split}
            \end{equation}
        \normalsize
        where the coefficients $a_{ij}$ and $b_{ij}$,  
        $i\in\{1,2,...,n\}$, $j\in\{1,2,...,r\}$  
        describe the magnitude of the \color{black} input %(3NOV2021 for R4-1) 
        \color{black} forced oscillation  of $\omega_{Fi}$ at the $j$th generator. %It should be noted that 
        %\color{blue} swing equation  (3NOV2021 for R4-1) \color{black} . %\color{green} (delete) Without forced oscillation, $\Xi_{ab}$ should be a matrix close to zero; otherwise, \color{blue} (Natural oscillation may causes non-zero in matrix) \color{black}
        When forced oscillations are contained in $X$, $\Xi_{a b}$ is a sparse matrix with large magnitudes only corresponding to the forced oscillation frequencies at the source generators. Thus, from the estimated $\Xi_{ab}$, we can identify the sources of forced oscillations. \color{black}It is worth noting that $a_{ij}$ and $b_{ij}$, or more specifically $\sqrt{a_{ij}^2+b_{ij}^2}$, describes the magnitude of the potential input forced oscillation ${u}_j$ at frequency $\omega_{Fi}$ (see (\ref{eq:inputFO})) that is different from the oscillation magnitudes observed from the state variables $[\Delta \bm{\delta}, \Delta \bm{\omega}]$ or other measurements (e.g., voltage magnitudes). \color{black}
        Besides, SINDy intrinsically can handle both linear and nonlinear dynamics (e.g. \cite{brunton2016sparse}). Since we consider sustained oscillations in ambient conditions rather than ring-down signals, the linearized power system model (\ref{eq:swingEqVect}) can be applied to describe the system dynamics. \color{black}%\color{blue} For the resonance case, \cite{foHuang} shows that the resonance response has a low rank structure, and it can be integrated with the low rank system matrix which is found by RPCA. In the proposed algorithm, objective of finding lowest rank matrix as RPCA is replaced with finding the matrix with highest sparsity. By regression the system dynamic, noise, and forced oscillation simultaneously, the system response is separated accordingly. \color{red} comment on why the method works for resonance case? \color{black}

        \color{black}
        %However, to implement SINDy and build up the feature library in 
        From (\ref{library}), it can be seen that the forced oscillation frequencies need to be estimated first in order to build up the feature library $\Theta(X)$. %To obtain the forced oscillation frequencies which are contained in the measurement matrix (\ref{eq:measurementMatrix}), 
        %the spectrum of measurements can be obtained through %the second-order Goertzel algorithm \cite{goertzel1958algorithm} 
        To this end, Fast Fourier Transform (FFT) is applied to the measurement matrix $X$ in (\ref{eq:measurementMatrix}). The z-score based peak detection method \cite{zscore} is used to capture the candidates for forced oscillation frequencies $\{\omega_F\}_i$ from the spectrum for each measurement.
        %(R1 SINDy 1)
        \color{black}
        %Fast Fourier Transform (DFT). 
        %Particularly,  the 
        %for DFT in this paper. 
        %Note that the maximum frequency at the spectrum is at most half of the measurement sampling frequency considering the Nyquist–Shannon sampling theorem.
        %Besides, the z-score based peak detection method \cite{zscore} is used to capture the peak frequencies $\{\omega_F\}_i$ for each measurement. 
        In brief, z-score is an index that describes the relationship of a measurement to the mean of a group of values. A z-score of $1.0$ means that the measurement's standard deviation is 1 away from the group mean. 
        The statistical dispersion gives a good indication for the anomaly such as the peak in the spectrum. 
        Once peak frequencies are identified for all measurements, we collect $n$ final candidate frequencies by  $\Omega_F=\bigcap^{2r}_{i=1}\{\omega_F\}_i$. 
        It should be noted that $\Omega_F$ may contain frequencies of both natural oscillations and forced oscillations.  %Next, we need to solve $\Xi$, from which the forced oscillation sources can be identified. 
        %\color{blue} move the measurement matrix after model as (R5-1) 
        \color{black}
        
        \color{black} Once the feature library is built, we need to solve $\Xi$. \color{black} %To solve $\Xi$,
        A sequential thresholded least square fitting sparse regression process developed in \cite{brunton2016discovering} is applied in this paper,  
        which starts with an initial estimation $\Xi^{0}$ from 
        the least square fit:
        
        \vspace{-3ex}
        \small
                \begin{equation}
                    \begin{split}
                        \Xi^{0} = {\Theta_{X}}^{\dagger} \dot{X} \vspace{-2ex}
                    \end{split}
                \end{equation}
        \normalsize
        where ${\Theta_{X}}^{\dagger} = ({\Theta_{X}}^{T} {\Theta_{X}})^{-1} {\Theta_{X}}^{T}$ %${\Theta_{X}}^{\dagger}$ 
        denotes the pseudo-inverse of ${\Theta_{X}}$. 
        To promote the sparsity of the matrix $\Xi$, a threshold $\lambda$ is introduced. 
        For the $j$th column $\bm{\xi_{j}^{0}}$ in $\Xi^0$, the indices for the values which are greater than $\lambda$  are recorded in the $j$th column $\bm{S_{j}^{new}}$ of the index matrix $S^{new}$.  
        Then the least square fit will perform only for the elements with indices in $\bm{S_{j}^{new}}$ in the new iterations to get $\bm{\xi_{j}^{new}}\left(\bm{ S_{j}^{new} }\right)$.
        The process is repeated until the index matrix $S$ remains the same between two iterations or the maximum iteration number is met. The pseudo code is presented in Algorithm 1. 
        
        It has been proved in \cite{zhang2019convergence} that SINDy algorithm approximates the local minimizer of the non-convex objective function with the zero norm penalty term as the following:
        
        \vspace{-3ex}
        \small
                \begin{equation} 
                    \begin{split}
                        F \left(\Xi \right) = \norm{\Theta_{X}\Xi- \dot{X}}_{2}^{2}+ \lambda^{2} \norm{\Xi}_{0}
                    \end{split}
                \end{equation}
        \normalsize
        To see that,  %SINDy algorithm, 
        after each iteration, the element  of $\Xi$ is either with a least square solution that is greater than $\lambda$ or zero, implying that the number of iterations for SINDy algorithm to \color{black} converge  \color{black} is finite and at most equals to the cardinality of $\Xi^{0}$, i.e., $card(\Xi^{0})$.
        Additionally, in the application of power systems, some elements of  $\Xi$ are known to be zero as derived from power system physical model (see (\ref{Xi_DCbias})-(\ref{Xi_ab})), 
        those parts are enforced to be zero during the regression process.
        
        \small
        \iffalse\fi
                \SetKwInput{kwInit}{Initialize}
                \SetKwInput{kwReturn}{Return}
                \SetKwInput{kwInput}{Input}
                \SetKwInput{kwOutput}{Output}
                \begin{algorithm}
                    \kwInput{$\Theta_{X}, \dot{X}, \lambda$}
                    \kwOutput{$\Xi^{new}$}
                    \kwInit{$\left\{\begin{array}{l} iter \gets 1 \\
                            \Xi^{old} \gets \Xi^{0} = {\Theta_{X}}^{\dagger} \dot{X} \\
                            S_{}^{old} \gets 0_{2r \times \left(1+2r+2n \right)}
                                \end{array} \right.$}
                    \While{$iter < card(\Xi^{0}) $}{
                        \For{$j \gets 1$ to $2r$} 
                        {
                            $\bm{S_{j}^{new}}=\left\{\begin{array}{ll}
                                1, & \text { if }\left\{\bm{\xi_{j}^{old}}\right\} \geq \lambda \\
                                0, & \text { if }\left\{\bm{\xi_{j}^{old}}\right\}<\lambda
                                \end{array}\right.$
                                
                            $\bm{\xi_{j}^{new}} \left(\bm{S_{j}^{new}}\right)=\Theta\left(:, \bm{S_{j}^{new}}\right)^{\dagger}\dot X(:,j)$ %\bm{\dot{x}_{j}};$
                        }
                        
                        \eIf {$S^{new} \neq S^{old}$}
                        {
                            $\begin{array}{ll} S^{old} \gets S^{new}; \\
                            iter++; \end{array}$
                        }
                        {
                            break; 
                        }
                    }
                    return
                    \caption{The Sequential Thresholded Least-Square Sparse  Regression}
                \end{algorithm}
        \normalsize

        If forced oscillations are detected, for example, from modal oscillation component properties and envelope shapes by \cite{ForcedOscillationDetection, tang2019periodogram,wang2015data}, their sources can be further located from index $\zeta$ which can be calculated from the estimated $\Xi_{ab}$: %by extending (\ref{eq:inputFO}) as:
            \begin{eqnarray} \label{eq:indexMatrix}
                \begin{split}
                %\zeta_{ij}=a_{ij}^{2}+b_{ij}^{2}\qquad i\in\{1,...,n\}, j\in\{1,...,r\}\\  
                \zeta_{ij} = a_{ij}^{2}+b_{ij}^{2} \quad %=  %\zeta_{ij} \cos^2 (\varphi_{ij}) &+ \zeta_{ij} \sin^2 (\varphi_{ij}) \\ 
                i\in\{1,...,n\}, j\in\{1,...,r\}
                \end{split}
            \end{eqnarray} 
        \color{black}
        %It is obvious that 
        \color{black}
        From (\ref{eq:inputFO})\color{black}, it can be seen that $\zeta_{ij}$ is the \color{black} estimated \color{black} magnitude squared of the \color{black} injected forced \color{black} oscillation of frequency $\omega_{F_i}$ at generator $j$.
        \color{black}
        Therefore, the forced oscillation frequencies $\omega_{F_{i}}$ and the source generators $j$ can be identified from the peak values in $\{\zeta\}$. Due to the sparsity of $\zeta$ matrix, outlier selection algorithms can be applied to pinpoint the peaks. An outlier is defined as a value that is more than three scaled Median Absolute Deviations (MAD)  away from the median \cite{MAD2013}. In particular, function \say{isoutlier} in Matlab\,\textsuperscript{\tiny\textregistered} R2019 %\color{red}give the version and add copyright mark\color{black} 
        can be applied to detect the forced oscillation frequencies and their sources.
        %the outlier is selected as a value that is more than three scaled median absolute deviations (MAD) away from the median which is immune to the sample size \cite{MAD2013}.
        %\color{blue}
        %Due to the high sparsity in the $\zeta$ matrix, the peak selection process can be considered as finding the outlier in $\zeta$ matrix.
        %The outlier is selected as a value that is more than three scaled median absolute deviations (MAD) away from the median which is immune to the sample size \cite{MAD2013}. 
        %In particular, function \say{isoutlier} in Matlab can be applied to detect the forced oscillation frequencies and their sources.
        %\color{black} 
        
\begin{figure}[!ht]
        \centering
                \includegraphics[width=3in]%height= 2 in ]
                {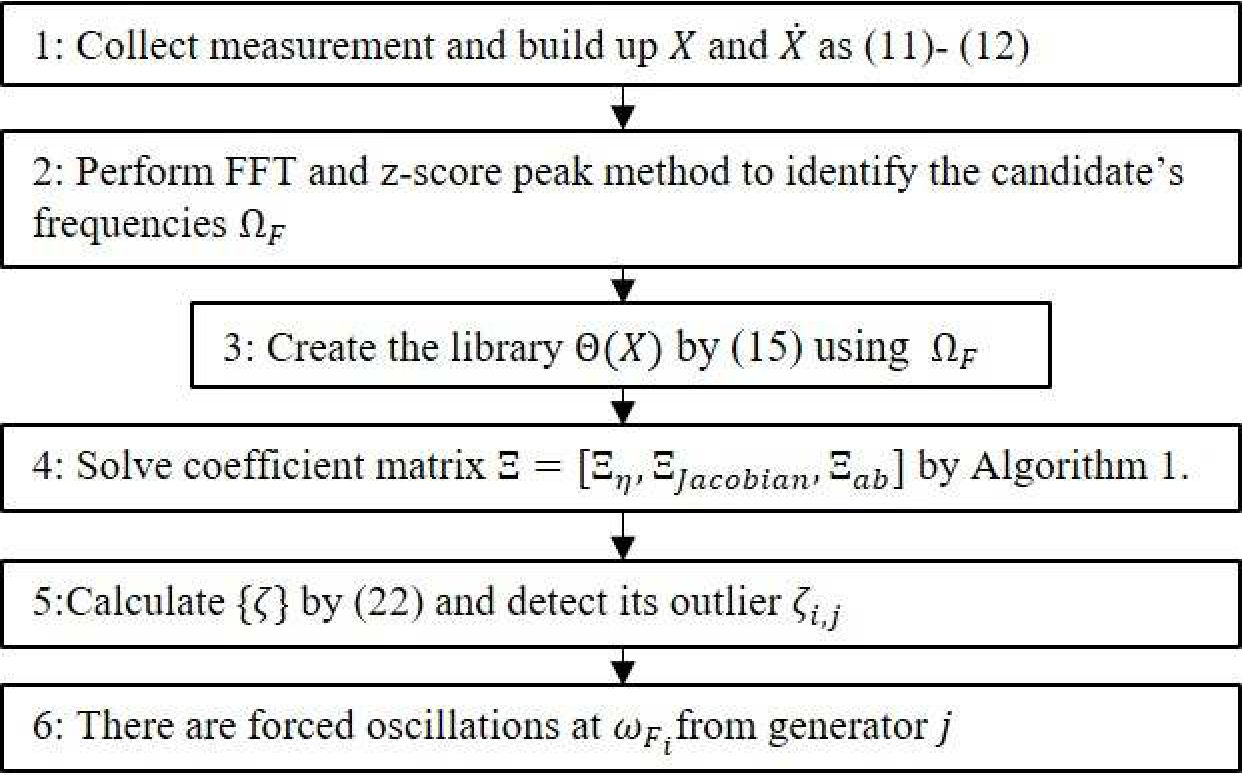}
                \caption{The flow chart of the proposed algorithm of locating forced oscillations.} %\color{red} check eqn number before finalization \color{blue} update the flow chart\color{black}} 
                \label{fig:flowChart}
        \end{figure}

        Assuming a forced oscillation event is detected, 
        the flow chart of the proposed algorithm in locating forced oscillations is summarized in Fig.\,\ref{fig:flowChart}, followed by some important remarks. 
        
        Remarks:\\
        \noindent$\bullet$ In Step 1, $m=1200$ is used for most of the cases in the library\cite{websiteForcedOsi} %case studies of this paper, 
        i.e., 40s PMU data with a sampling frequency of 30 Hz, except for Cases 4 and 5 in the actual oscillation events in ISO-NE system where 80s PMU data is used due to weak oscillation magnitudes. %because the forced oscillation magnitude is too small. %\color{red} please confirm \color{red} How about Case 4?\color{black} 
        The applied sample size ensures that the forced oscillation frequencies can be captured accurately. %and the estimation accuracy for $\Xi$ is sufficiently good.  %which yields good estimation accuracy for $\Xi$. 
        The computational time of the proposed algorithm is %negligible,
        approximately $3$ seconds for all cases using Matlab\,\textsuperscript{\tiny\textregistered} R2019 software on Intel i7-7700HQ CPU with 16GB RAM computer,
        %The relatively small sample size and trivial computational cost 
        which demonstrates a good feasibility of the proposed algorithm in online applications.     
        
        %\color{blue}
        \noindent$\bullet$ The proposed method requests the derivatives of frequencies. Although Rate Of Change Of Frequency (ROCOF) is standard measured data \color{black}\cite{IEEEc37d118}\color{black}, %\color{green} (delete) according to IEEE Standard for Synchrophasor Data Transfer for Power Systems \cite{IEEEc37d118} \color{black}, 
        it may be unavailable or \color{black}inaccurate \color{black} in practice. 
        %or \color{blue} inaccurate \color{black} in practice due to undesirable components like harmonic frequency, and noise\cite{IEEEc37d118}. 
        A numerical finite difference method, specifically, the two-points forward difference approximation method  \cite{lindfield2018numerical} is applied to estimate the derivatives of frequencies in the simulation studies of this paper if they are not available. 
        %The proposed method requests the derivatives of frequencies. Although Rate Of Change Of Frequency (ROCOF) is standard measured data, 
        %it may be unreliable in practice due to large frequency measurement error from undesirable components like harmonic frequency, and noise \cite{IEEEc37d118}. 
        \color{black}
        In practice, the frequency measurement filter proposed in \cite{pierre2019design} or the frequency estimation algorithm using the point-on-wave model in\cite{7741706} may further enhance the measurement accuracy and improve the performance of the proposed method. \color{black} %An advanced frequency estimation algorithm with the point-on-wave model algorithm from \cite{7741706}, and the frequency measurement filter design which is introduced in \cite{pierre2019design} is needed for real-world deployment.
        
        %Besides, more advanced methods like the point-on-wave model algorithm and an advanced filter with interpolation can be applied 
        %which yields sufficiently good accuracy. Besides, 
        % tested for proposed algorithm which includes two or three-points forward and center difference approximation with good results. Since the reporting rate is relatively high with $30$ frames per second. The two-points forward difference approximation is used.
        %\color{blue}
        %An advanced frequency estimation algorithm with the point-on-wave model algorithm from \cite{7741706}, and the frequency measurement filter design which is introduced in \cite{pierre2019design} is needed for real-world deployment. % (R1 Data Driven Method 2)
        %\color{black}
        
        \color{black}
        \noindent$\bullet$ In this paper, FFT is applied to identify forced oscillation frequency candidates. Goertzel algorithm and Welch's method can also be applied as alternatives for a higher resolution spectrum.  
        %alternative, which may be more efficient in identifying forced oscillation frequencies in a particular band. Besides, Welch's method and other advanced alternatives may also be applied for a higher resolution spectrum. 
        Since FFT, Goertzel and Welch algorithms can provide good candidate sets in all the numerical studies presented in Section \ref{section:numerical study}-\ref{section:real case testing}, we only present the results using FFT. 
        \color{black}

        \noindent $\bullet$ If the forced oscillation sources are outside the study area, the border bus treated as an aggregation of the external area may be identified as the source by the proposed algorithm as shown in Section \ref{section:FurtherValidationsReal}. %will treat the border bus as generator bus as an aggregation of the system. 
        Further effort is needed to extend the method to investigate the situations in which the forced oscillation sources are not generators. \color{black}
        
        \color{black}
        \noindent $\bullet$ In all cases presented in Section \ref{section:numerical study}-\ref{section:real case testing}, $\lambda$ is selected as $10^{-6}$ that gives an expected sparsity level for $\Xi_{ab}$ ($>50\%$, see (\ref{eq:Xi})-(\ref{Xi_ab})). In practice, the value of $\lambda$ may need to be adjusted in different systems and can be determined considering the expected sparsity level of $\Xi_{ab}$. Besides, 
        %However, the value of $\lambda$ varies due to the strength of the oscillation being different case by case in the actual world practice.
        %Since the sparsity of $\Xi$ matrix is beyond $50\%$ as in (\ref{eq:systemSINDy}), the sparsity promoting parameter $\lambda$ can be adjusted dynamically by ensuring the result $\Xi$ matrix has more than $50\%$ zero elements.
        the sparse promoting process for the thresholded Least absolute Shrinkage and Selection Operator (thresholded LASSO) \cite{kulkarni2020sparse} can also be used to promote the non-zero coefficients in the proposed algorithm without the need of selecting $\lambda$.
        \color{black}

    \section{\uppercase{testing in simulated cases}}\label{section:numerical study}

        In this section, the performance of the proposed algorithm of locating forced oscillations is tested in all 24 cases available in the IEEE Task Force test cases library \cite{realCasePaper, websiteForcedOsi}. %\cite{realCasePaper, websiteForcedOsi}. 
         \color{black} In all cases, all natural modes are reasonably damped with damping ratios higher than $5\%$.  \color{black} 
         \color{black}
        The test results of the proposed algorithm  in the IEEE 68-bus system and the WECC 240-bus system are also presented.
        \color{black}
        It will be shown that the proposed method can successfully detect forced oscillations sources even when the system is under resonance condition (i.e., the forcing frequency overlaps with a natural mode). %while the natural model is poorly damped. %and the forcing strength is low. 
        
        \color{black}

        %%%%%%%%%%%%%%%%%%%%%%%%%%%%%%%%%%%%%%%%%%%%%%%%%%%%%%%%%%%%%%%%%%%%%%%%%%%%%%%%%%%%%%%
        \vspace{-0.15in}
        \subsection{Case A: One Forced Signal, Resonance with an Inter-Area Mode} %Oscillation at a Resonant Frequency} 
        
        %\color{blue}
        One challenging situation for any forced oscillation locating method is when resonance occurs such that large magnitude oscillations are observed in multiple locations. In light of this, we test the performance of the proposed algorithm under resonance conditions. Particularly, we take Case $F3$ in the forced oscillation simulated test case library \cite{realCasePaper} as an example, which is also available online \cite{websiteForcedOsi}. The performance of the proposed method in all 18 simulated cases is also presented in Table \ref{tab:SINDyRPCAcomparison}. %at the end of Section \ref{section:two source case}. 
        %\cite{maslennikov2016test} provide a simulation data library which is designed  to test the performance of forced oscillation locating method in different scenario which include Forced Oscillation resonance with the  local and inter-area mode. 
        The library dataset is produced by TSAT software from simulation of WECC 179-bus system. 
        
        %\color{red} Introduce Case $F3$ \color{black}  \color{red} I think the time-domain trajs need to be presented to show that the source does not correspond to the largest magnitude oscillation
        
\begin{figure}[!ht]
                \centering
                \includegraphics[width=3.6 in]{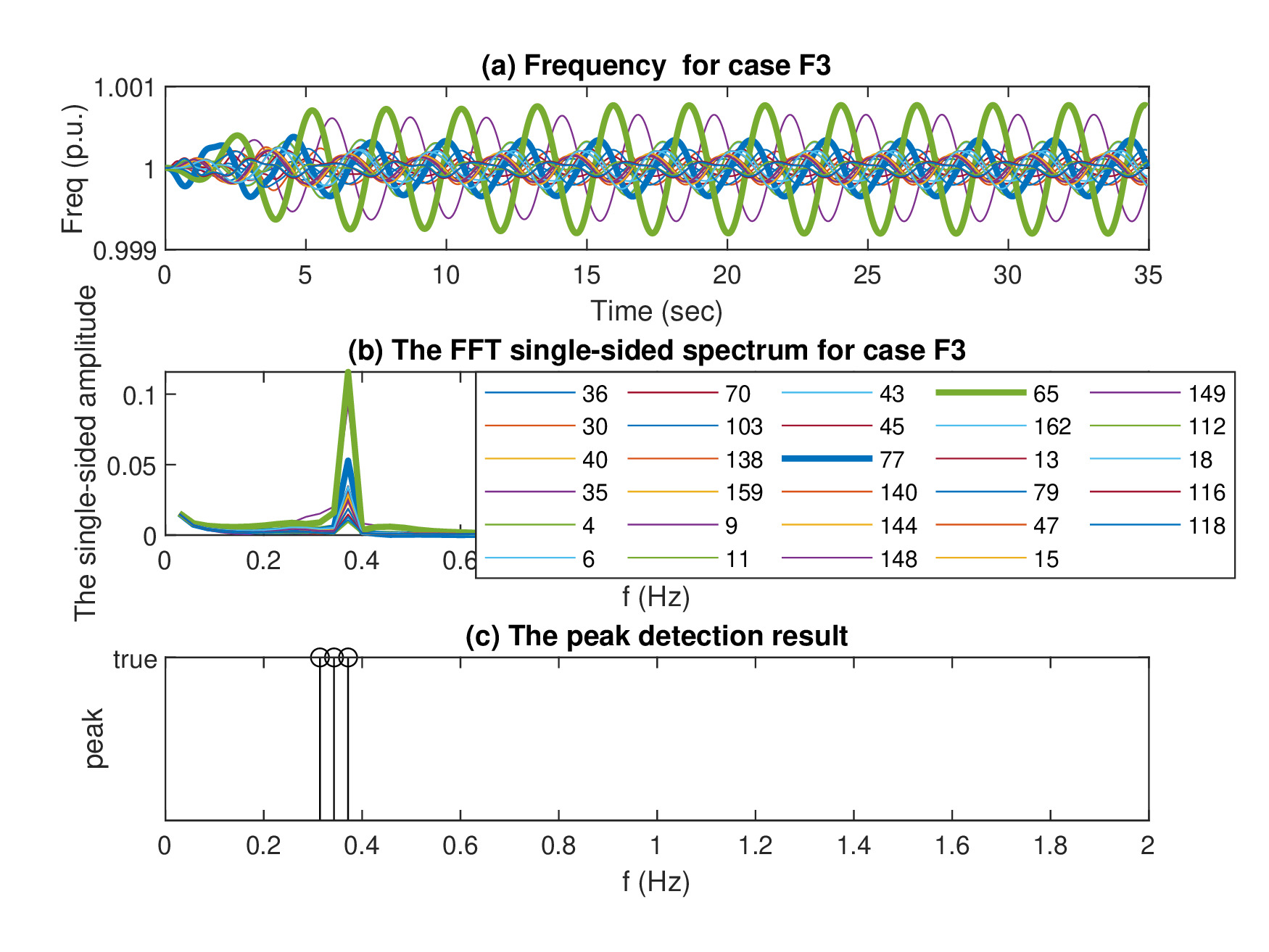}
                \caption{(a). The trajectories of rotor frequencies; \color{black}(b) The single-sided spectrum from FFT;\color{black} (c) The peak frequencies detected for Case A (Case $F3$ in \cite{websiteForcedOsi}) . } 
                \label{fig:F3_fft}
                \vspace{-0.15in}
            \end{figure}

        %\color{blue} %Case $F3$ is a simulated forced oscillation event in which a resonance with inter-area 0.37Hz mode appears. 
        In Case $F3$, forced oscillation is injected into the excitation system as a sinusoidal signal at the generator at \color{black} bus \color{black} 77 with a frequency 0.37Hz to cause resonance with the inter-area 0.37 Hz mode. More details can be found in \cite{websiteForcedOsi}. 
        %phenomenon between the forced oscillation appeared at generator excitation system in generator which connect to bus $77$ and the $0.37$ Hz inter-area mode. 
        The trajectories of frequencies are presented in Fig.\,\ref{fig:F3_fft} (a). The largest oscillation amplitude in frequency is observed at the generator at bus $65$ (marked by bold green) rather than the actual source, i.e., the generator at bus $77$ (marked by bold blue). %\color{red} can you mark on Fig. 4 (a), the trajectories for the two generators respectively. If it is too small, you may draw a separate figure. \color{blue} (A bold line is used for the source trajectories) \color{black}
        %that is far away from the actual oscillation source  
        %The presence of resonance phenomena will cause the largest oscillations region is far away from the source region \cite{InterAreaResonance}, especially the forced oscillation frequency is near the inter-area modes.
        %As seen from the single-sided amplitude spectrum in Fig.\,\ref{fig:F3_fft} (b), the largest amplitude at forced oscillation frequency is at the distant generator bus $65$ (color green) instead of the source (color blue). %which indicates a resonance condition occurred.
        %\color{red} not sure if that "indicates" \color{black}
        
        Following the proposed algorithm, %The proposed algorithm uses 
        $40$s, $30$Hz emulated rotor angles and frequencies %obtained from PMU measurements of all generators to 
        are used to build up the measurement matrix $X$ and $\dot X$. The finite difference method is applied to estimate $\dot X$. In Step 2. the FFT and the z-score peak detection method are applied to detect the peak oscillation frequencies. 
        As seen from the single-sided amplitude spectrum in Fig.\,\ref{fig:F3_fft} (b), natural frequencies and forced oscillation frequencies are mixed. 
        After removing $26$ duplication, $3$ candidate frequencies are collected in $\Omega_F$ as presented in Fig.\,\ref{fig:F3_fft}. (c). Particularly, $0.37$ Hz, the true forced oscillation frequency is identified, which overlaps with a natural inter-area mode. %\color{blue} The system has a natural mode at $0.37$Hz, the true forced oscillation frequency is overlapped with the natural mode and hiding among the candidate frequencies.  %\color{red} can you comment on which freq is forced and which is natural?
        \color{black}
          
        Next, the feature library $\Theta(X)$ is built in Step 3 using 3 candidate frequencies. The coefficient matrix $\Xi=\left[
        \bm{\Xi_{\eta}},\Xi_{\text {Jacobian}}, \Xi_{a b}\right]$ is further obtained in Step 4. Particularly, $\Xi_{ab}$ corresponding to the forced oscillation terms is shown in Fig.\,\ref{fig:F3_result} (a), from which it can be evidently seen that there is one peak term at $0.37$ Hz. %\color{red} i think it is better to use either $0.366$ or $0.37$, be consistent\color{black}
        %\color{red} I cannot see two peaks. I also don't see 0,7 Hz and 0.5Hz \color{black}
        
        Following the proposed algorithm, the indexes $\{\zeta\}$ are calculated and the outliers are detected, as shown in Fig.\,\ref{fig:F3_result} (b)-(c). It is determined by the proposed algorithm that there is one oscillation source, i.e., the generator at bus 77, at a frequency of $0.37$ Hz, which well agrees with the description of the source for this case in the test library.  In addition, it takes only $3.07$ seconds for locating, showing a high computational efficiency of the proposed method.

        %demonstrating that the proposed algorithm pinpoint forced oscillation sources accurately and efficiently. 
        %The location result from proposed algorithm for case $F3$ is shown in Fig.\,\ref{fig:F3_result}. It is determined by the proposed algorithm that there are one forced oscillation source which is in bus $77$ at  $0.37$ Hz.  
        %The result is agreeing with the description of the case $F3$. However, it is less sparesly compare to previous IEEE 68 bus simulation, due to the estimation error is increasing as the sampling frequency decreased from $60$ Hz to $30$ Hz. \color{red} many typos\color{black}
        \iffalse
            \begin{figure}[!ht]
                \centering
                \includegraphics[width=3.5 in, height= 2.2 in]{F3_result2_14apr2021.jpg}
                \caption{$\zeta$ (left) and outlier (right) result for Case F3}
                \label{fig:F3_result}
                \vspace{-0.1in}
            \end{figure}
        \fi    
        \begin{figure} [!ht]
        \vspace{-0.0in}
            \centering
            \vspace{-0.0in}
            \includegraphics[width=3.5in]{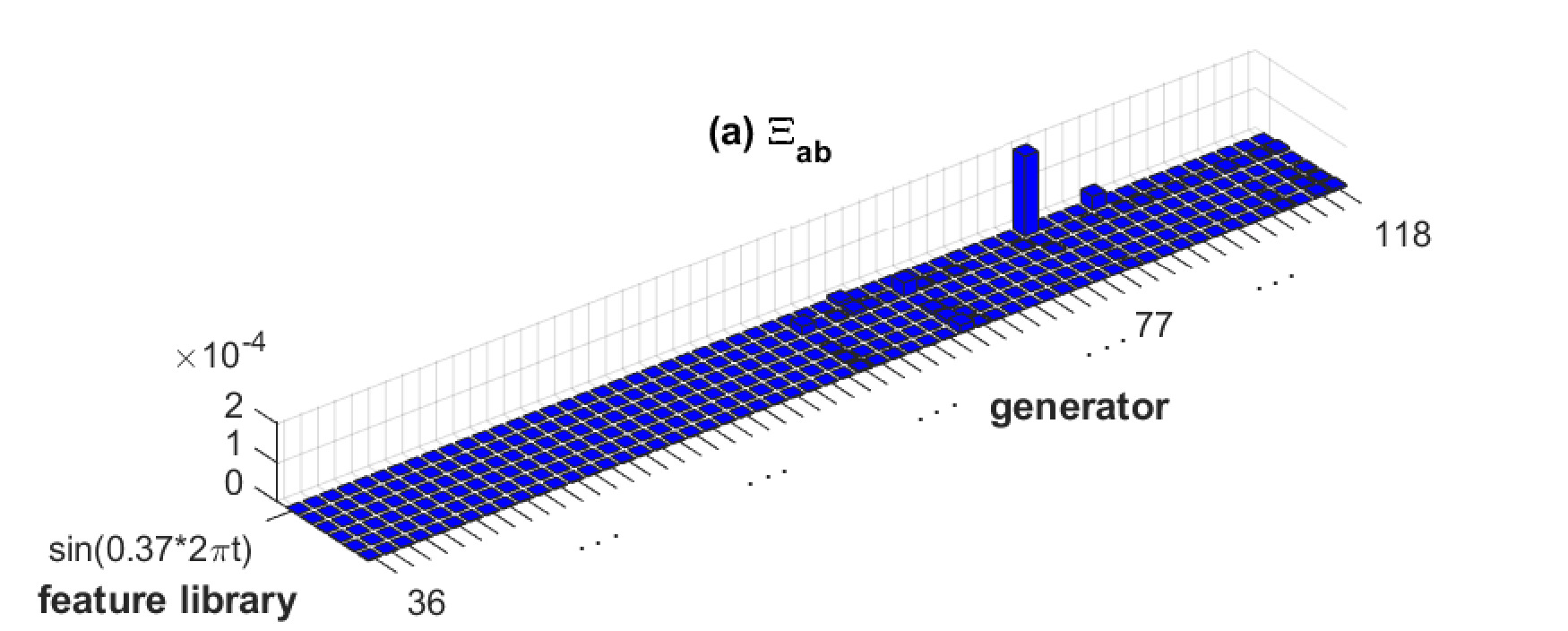}
            \vspace{-0.0in}
            \includegraphics[width=3.5 in]{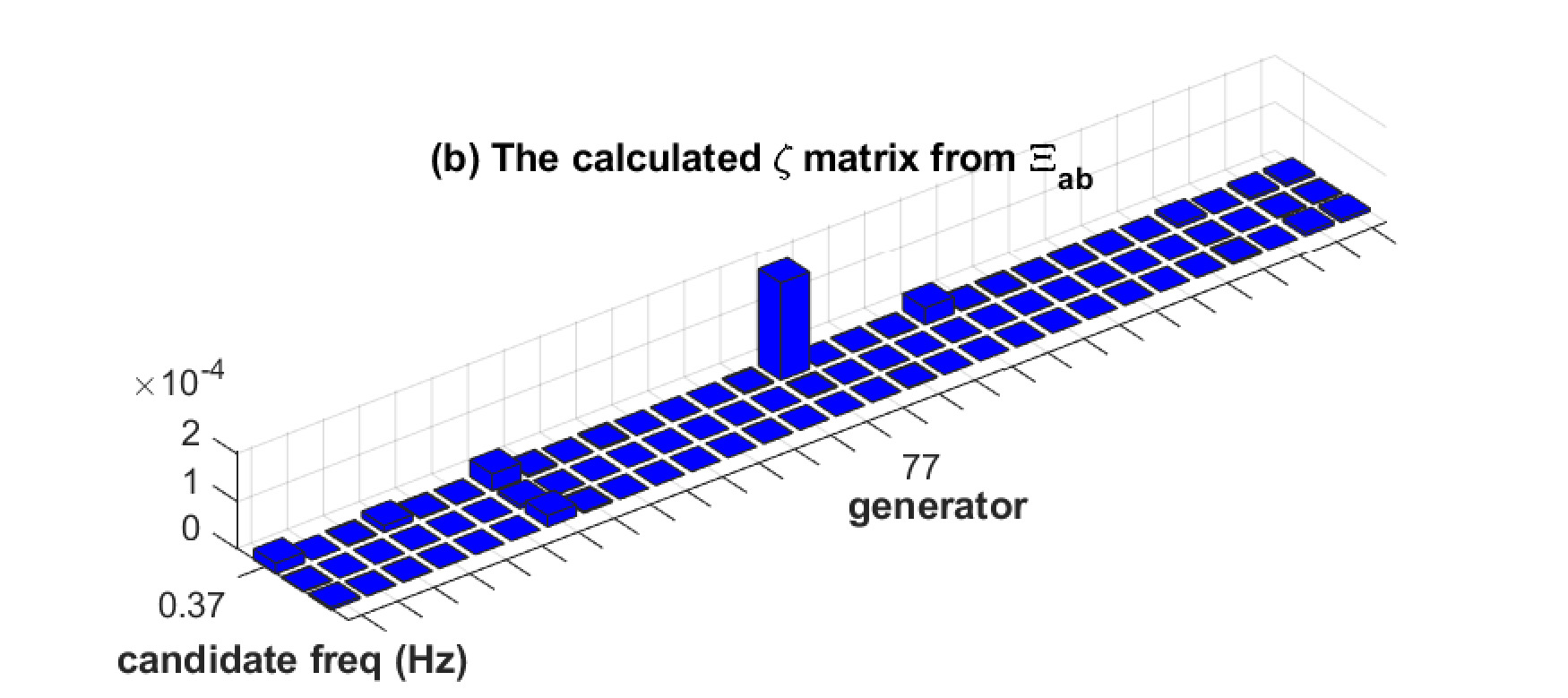}
            \vspace{-0.0in}
            \includegraphics[width=3.5 in]{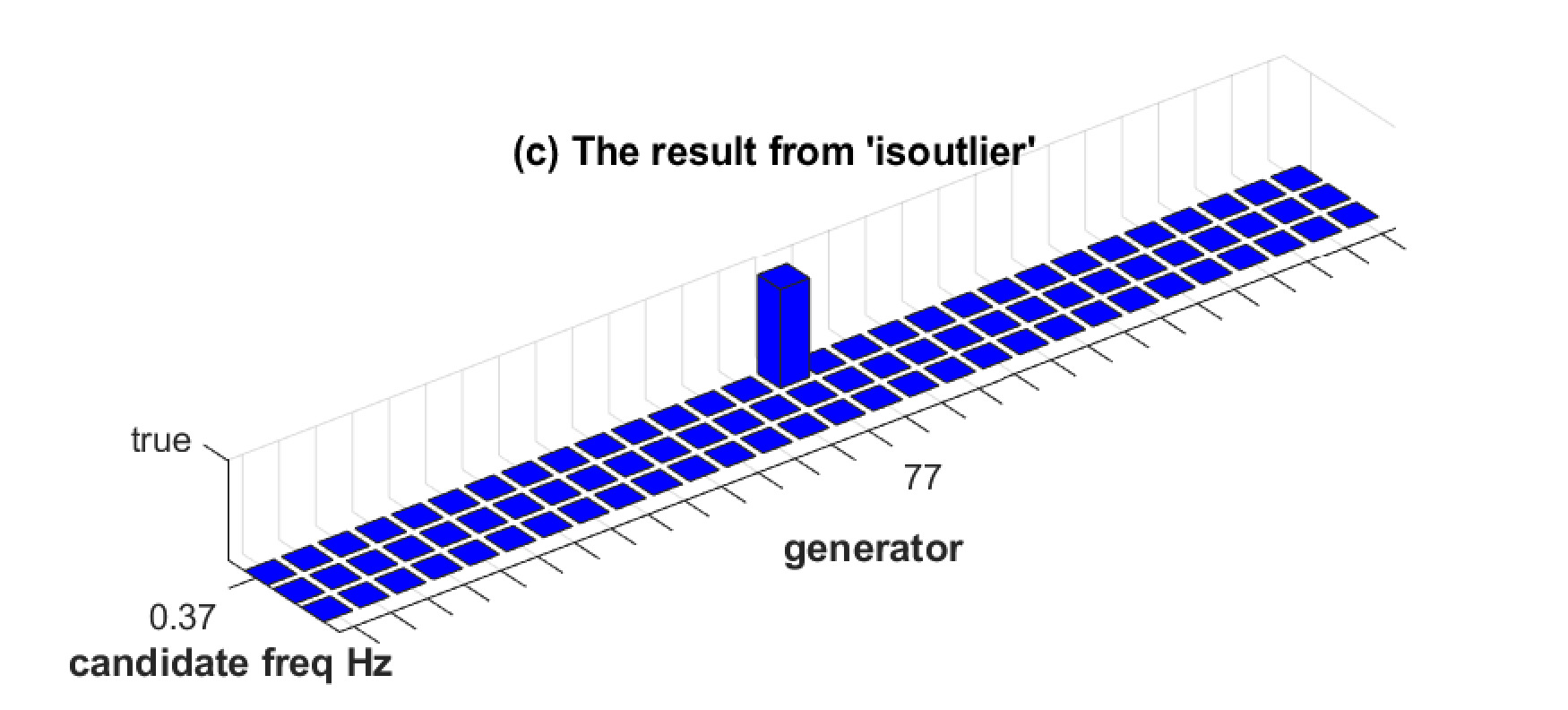}
            \caption{(a) $\Xi_{ab}$, (b) $\zeta$, and (c) outliers of $\zeta$ for Case A (Case $F3$ in \cite{websiteForcedOsi}).}
            \label{fig:F3_result}
        \vspace{-0.2in}
        \end{figure}

        \color{black}

        %%%%%%%%%%%%%%%%%%%%%%%%%%%%%%%%%%%%%%%%%%%%%%%%%%%%%%%%%%%%%%%%%%%%%%%%%%%%
        %\vspace{-0.15in}
        \subsection{Case B: Under a Rectangular Forced Oscillation} 
        
\begin{figure}[!ht]
            \centering
            \includegraphics[width=3.5 in]{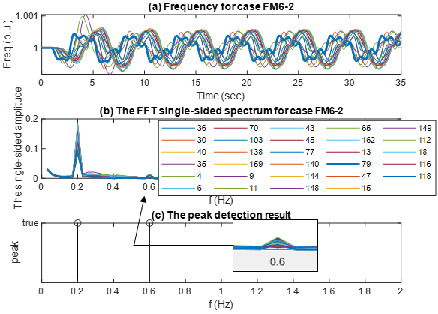}
            \caption{(a). The trajectories of rotor frequencies;\color{black}(b) The single-sided spectrum from FFT;\color{black} \color{black}(c) The peak frequencies detected for Case B (Case $FM6-2$ in \cite{websiteForcedOsi})\color{black}.} \vspace{-0.2in}
            \label{fig:FM6_2_fft}
            %\vspace{-0.15in}
        \end{figure}

\begin{figure} [!ht]
        %\vspace{-0.0in}
            \centering
            \vspace{-0.0in}
            \includegraphics[width=3.6 in]{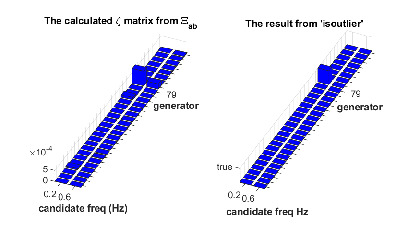}
            \caption{$\zeta$ (left) and outlier (right) result for Case B (Case $FM6-2$ in \cite{websiteForcedOsi}).} \vspace{-0.2in}
            \label{fig:FM6_2_result}
        %\vspace{-0.2in}
        \end{figure}

        %\color{blue}
        In Case $FM6-2$ %from the test case library 
        \cite{websiteForcedOsi}, forced oscillation is injected into governor as a rectangular and symmetric signal %sinusoidal signal 
        at the generator at bus 79.  The forced signal creates the spectrum of odd harmonics 0.2, 0.6, 1.0, 1.4Hz, etc. with the lowest frequency 0.2Hz \cite{websiteForcedOsi}. %with fundamental frequency as 0.2Hz \cite{websiteForcedOsi}. 
        The trajectories of frequencies are presented in Fig.\,\ref{fig:FM6_2_fft} (a), showing that the signal of the source (marked in bold blue) does not have the largest oscillation magnitude. Similarly, $40$s, $30$Hz emulated rotor angles and frequencies are used by the proposed algorithm. The FFT and the z-score peak detection method for the trajectories are presented in Fig.\,\ref{fig:FM6_2_fft} (b)-(c). 
        After removing $26$ duplication, $2$ candidate frequencies are collected in $\Omega_F$ as presented in Fig.\,\ref{fig:FM6_2_fft}. (c). %\color{red} is it a resonance case? \color{blue} it is not a resonance case \color{black}
        The indexes $\{\zeta\}$ are calculated and the outliers are detected. As shown in \color{black} Fig.\,\ref{fig:FM6_2_result} (left) and (right) \color{black}, the generator at bus 79 is detected to be the source of the oscillation, %, at a frequency of $0.2$ Hz, 
        which agrees with the description.  
        The processing time for locating is $3.04$ seconds.
        
\begin{figure}[!ht]
            \centering
            \includegraphics[width=3.6 in]{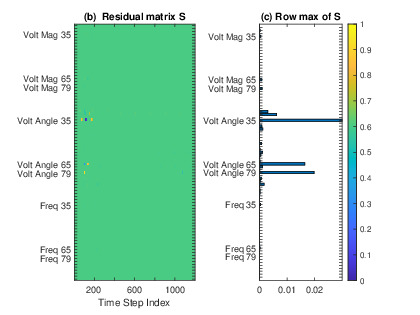}
            \caption{Visualization of the sparse residual matrix $S$ by the RPCA method (left) and the maximum absolute value in each row of $S$ (right) Case B (Case $FM6-2$ in \cite{websiteForcedOsi}).}
            \label{fig:FM6_2_RPCA_result}
            \vspace{-0.1in}
        \end{figure}

        For comparison, the RPCA method proposed in \cite{foHuang} was implemented using the same measurements. Leveraging on the sparsity of the forced oscillation sources along with the low-rank nature of rotor dynamics, the RPCA method computes the sparse and low-rank components
        of the measurement matrix and detects the forced oscillation sources by finding the largest absolute value in the sparse residual matrix $S$. 
        %we also used 
        %\color{red}Is the the same measurement used by SINDy? If so, you should mention what measurements are used at the beginning of this subsection and mention here again that the same measurements are used.  \color{blue} 
        %The 40 seconds combination measurements of terminal bus voltage magnitude, voltage angle, and frequencies data from the same measurements from Case $F3$  is used to build up the measurement matrix for RPCA locating method \cite{foHuang}.
        % \color{black}
        The %low rank approximation matrix $L$ and 
        sparse residual matrix $S$ and its row maximum absolute values calculated by the RPCA method  are shown in Fig.\,\ref{fig:FM6_2_RPCA_result}, from which it can be seen that the maximum absolute value of $S$ has a row index corresponding to bus 35.  
        %\color{blue} The comparison between RPCA and SINDy result is shown in Fig.\,\ref{fig:FM6_2_SINDyRPCA_result}. \color{black}  %correspond to bus 35. Therefore, 
        %\color{red} one sentence explain how the source is obtained from the two matrices and/or how RPCA works. 
        %\color{blue} 
        %As discussed in Section \ref{section:dynamic model}, by leveraging the sparsity of the forced oscillation sources along with the low-rank nature of rotor dynamics, the forced oscillation source detection is treated as separating the data matrix as a set of sparse residual matrix $S$ and low rank matrix $L$ with same dimension by using Robust Principal Component Analysis (RPCA).
        %The row index contains the maximum value from $S$ is deemed as the source of the forced oscillation.
        %\color{black} 
        %In this case, 
        Thus, the RPCA method regards the generator at bus $35$ to be the source, failing to pinpoint the right source, i.e., the generator at bus $79$. %\color{red} Can you explain the potential reason for the failure? Also, which previous methods mentioned in the introduction may not handle rectangular wave?
        %\color{blue}
        %for the failure might be that some of the low rank resonances leak from the low rank matrix $L$ to residual matrix $S$.
        %(Paper before 2017 don't use the test library. Paper after 2017 mainly use case F1 and some self modification to the F1 to validate. 
        %I do think all of them works, because no matter how complex the frequency will be, there will be one fundamental frequency. You can estimate the generator parameter, transfer function, or magnitude, phase response just for the fundamental frequency, and anomie will appear as response is drafting away from the normal value. 
        %But I am not sure. RPCA should work theoretically speaking, but after test it is not for some cases.)
        %\color{black}
        %As shown in Fig.\,\ref{fig:FM6_2_RPCA_result} (c), the index of the maximum from the sparse residual matrix detect generator at bus $35$ as the source of the forced oscillations. 
        %More comparison and discussion are in Section \ref{section:two source case}
        %\color{black}
        
        %\color{red} why it is more than the measurements used in SINDy? Should the same number of measurements be used? \color{blue} (In the Xie's paper, they are using 400 measurements. But I do change the window size as same as SINDy with 40 second data, both F3 and FM3 case is correct! However, FM6-2 is still wrong, so I add it into paper.) \color{black} 

        %\begin{figure}[!ht]
        %    \centering
        %    \includegraphics[width=3 in, %height= 2 %in]{figures/FM6_2_SINDyRPCA_19apr2021.jpg}
        %    \caption{RPCA result (left) and SINDy result (right) comparison for Case $FM6-2$}
        %    \label{fig:FM6_2_SINDyRPCA_result}
        %    \vspace{-0.1in}
        %\end{figure}

        %\color{red} I think we can keep only two figures in Fig. \ref{fig:FM6_2_RPCA_result}. Maybe only (b) and (c)? \color{blue} changed \color{black}
        
        %%%%%%%%%%%%%%%%%%%%%%%%%%%%%%%%%%%%%%%%%%%%%%%%%%%%%%%%%%%%%%%%%%%%%%%%%%%%
        \vspace{-0.15in}
        \subsection{Case C: Two Sources of Forced Signals, Resonance with Two Modes} 
        \label{section:two source case}
        
\begin{figure}[!ht]
            \centering
            \includegraphics[width=3.6 in]{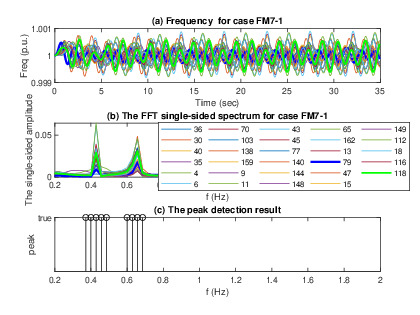}
            \caption{(a). The trajectories of rotor frequencies; \color{black} (b) The single-sided spectrum from FFT ; \color{black} (c) The peak frequencies detected for Case C (Case $FM7-1$ in \cite{websiteForcedOsi}).}
            \label{fig:FM7_1_fft}
            \vspace{-0.15in}
        \end{figure}

\begin{figure} [!ht]
        \vspace{-0.0in}
            \centering
            \includegraphics[width=3.6 in ]{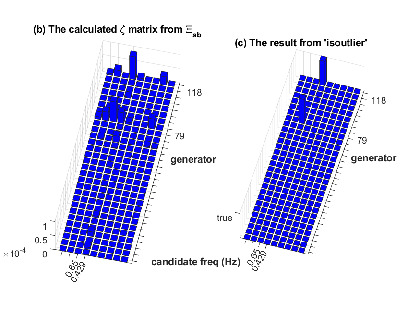}
            \caption{$\zeta$ (left) and outlier (right) result for Case C (Case $FM7-1$ in \cite{websiteForcedOsi}). }%\color{red} might save space if you can put them side by side \color{blue} I changed the plot\color{black}}
            \label{fig:FM7_1_result}
        %\vspace{-0.1in}
        \end{figure}

        %\color{blue}
        In Case $FM7-1$\cite{websiteForcedOsi}, two forced sinusoidal signals with fundamental frequencies $0.65$Hz and $0.43$Hz are injected into the governors of generators at bus $79$ and that at bus $118$, respectively, which are very close to two inter-area modes $0.667$ Hz and $0.445$ Hz. %as two sinusoidal signals with fundamental frequency as $0.65$Hz and $0.43$Hz, respectively \cite{websiteForcedOsi}. 
        The trajectories of frequencies are presented in Fig.\,\ref{fig:FM7_1_fft} (a), in which the source trajectories are marked in bold blue and light green. Again, the oscillations at the sources do not have the largest amplitudes among all frequency measurements, which is a typical sign of resonance \color{black} as discussed in \cite{Impedance}\cite{foHuang}. \color{black}
        %\color{blue}
        %As shown in Fig.\,\ref{fig:FM7_1_fft} (a) and (b), where the source measurement does not correspond to the most severe oscillation.
        %(R4 - 4) \color{black}
        %As shown in the FFT spectrum Fig.\,\ref{fig:FM6_2_fft} (b), the largest oscillation amplitude in each frequency is other than the actual source indicates a resonance condition occurred. 
        After removing $80$ duplication, $9$ candidate frequencies containing both forced oscillation frequencies and natural modes are collected in $\Omega_F$ as presented in Fig.\,\ref{fig:FM7_1_fft} (c). %Both forced oscillation frequencies ($0.429$ Hz and $0.65$ Hz) and natural oscillation frequencies ($0.445$ Hz and $0.667$ Hz ) are included in the identified candidate frequencies. \color{blue} 
        %Among the 9 candidates, the  true  forced  oscillation  frequency  are  identified as $0.429$ Hz and $0.65$ Hz, which are close to the natural inter-area modes $0.445$ Hz and $0.667$ Hz . %\color{red} Can you comment on if 9 candidate freqs include both forced osc freqs and natural modes? %\color{blue} %(if you zoom in, for some channels,  the spikes are not thin so the peak detection consider as some spikes within a big spikes. Actually, one of the objective for the SINDy is select the correct frequency) 
        \color{black}
        Again, the indexes $\{\zeta\}$ are calculated and the outliers are detected. As shown in Fig.\,\ref{fig:FM7_1_result}, the generators at bus 79 and bus 118 are pinpointed as the oscillation sources, %at frequencies of $0.65$Hz and $0.43$Hz, 
        which agree with the description from the test case. It is worth noting that there are no peaks/outliers detected at the natural oscillation frequencies, even though the forced oscillation frequencies are very close to the natural ones. %showing the capabilities of the proposed algorithm in handling resonance. 
        The processing time for locating the sources is $3.4$ seconds. 
        
\begin{figure}[!ht]
            \centering
            \includegraphics[width=3.6in ]{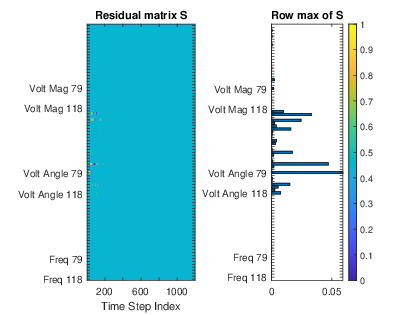}
            \caption{Visualization of the sparse residual matrix $S$ by the RPCA method (left) and the maximum absolute value in each row of $S$ (right) for Case C (Case $FM7-1$ in \cite{websiteForcedOsi}).}
            \label{fig:FM7_1_RPCA_result}
            \vspace{-0.1in}
        \end{figure}

        For comparison, the results of RPCA method using the same measurements are presented in Fig.\,\ref{fig:FM7_1_RPCA_result}. The row index of the maximum absolute value of the sparse residual matrix $S$ corresponds to the generator at bus 79. However, the other source, i.e, the generator at bus 118, is missed. Note that the RPCA method in \cite{foHuang} detects the forced oscillation source by observing the maximum absolute value of $S$ and thus is intrinsically unable to handle the cases with multiple forced oscillation sources. Moreover, the component corresponding to the true source, the generator at bus 118, does not even rank high in terms of absolute value as seen in Fig. \ref{fig:FM7_1_RPCA_result} (right). 
        %was also implemented. Low rank approximation matrix $L$ and sparse residual matrix $S$ for case $FM7-1$ are shown in Fig.\,\ref{fig:FM7_1_RPCA_result} (a)-(b). 
        %\color{blue}
        %The maximum index from the sparse residual matrix $S$ is used to detect the source of forced oscillations.
        %\color{black}
        %As shown in Fig.\,\ref{fig:FM7_1_RPCA_result} (c), the only index of the maximum from the sparse residual matrix detect generator at bus $79$ as the source of the forced oscillations. 
        %\color{black}
        %\begin{figure}[!ht]
        %    \centering
        %    \includegraphics[width=3 in, height= 2 in]{figures/FM7_1_SINDyRPCA2_19apr2021.jpg}
        %    \caption{RPCA result (left) and SINDy result (right) comparison for Case $FM7-1$}
        %    \label{fig:FM7_1_SINDyRPCA_result}
        %    \vspace{-0.1in}
        %\end{figure}
        \begin{figure}[!ht]
            \centering
            \includegraphics[width=3.6 in]{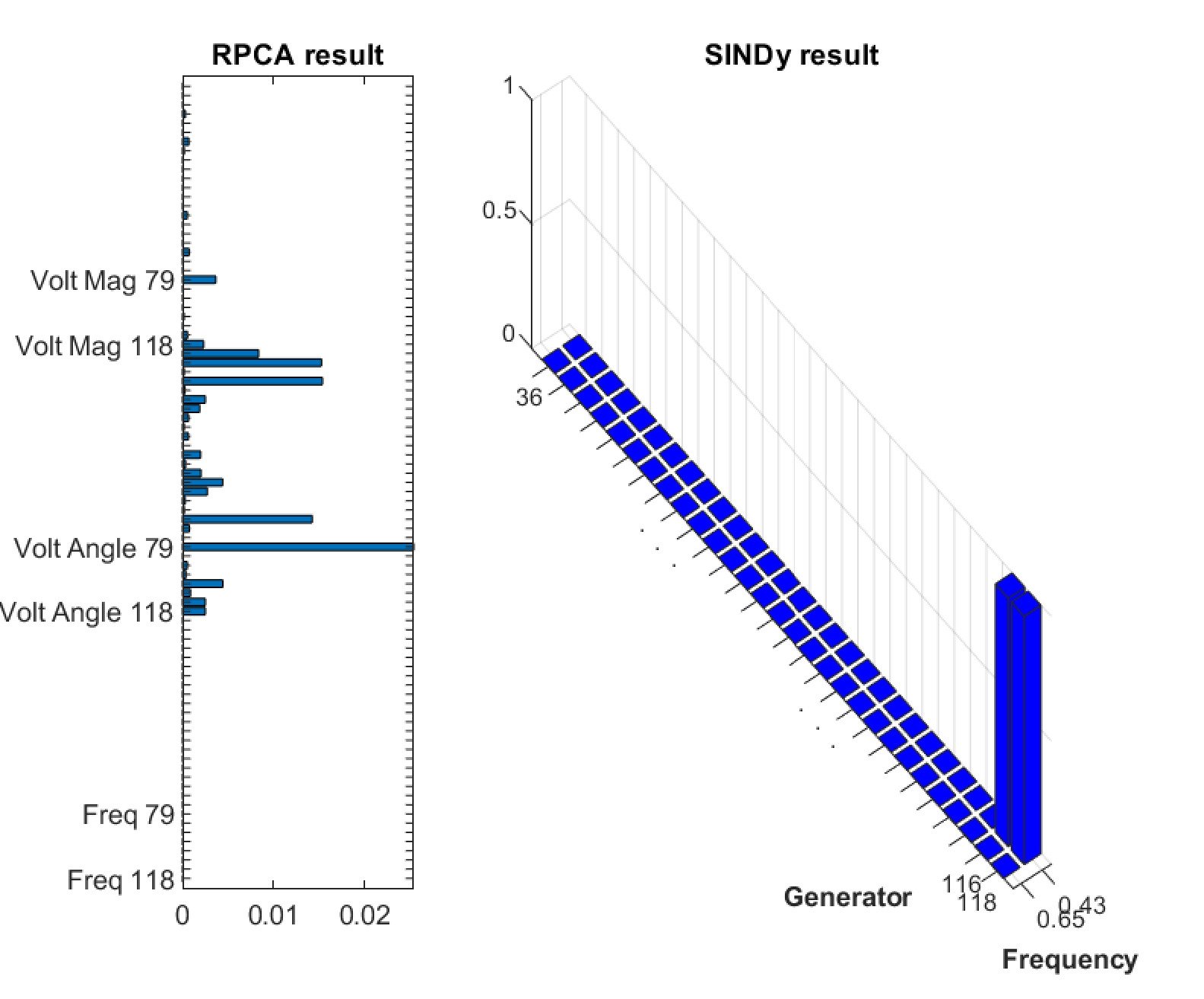}
            \caption{Visualization of the  RPCA result (left) and SINDy result (right) comparison for Case $F7-1$ in \cite{websiteForcedOsi}.}
            \label{fig:F7_1_SINDyRPCA_result}
            %\vspace{-0.1in}
        \end{figure}
        
        In addition, the proposed algorithm and the RPCA method \cite{ForcedOscillationDetection} are tested for all $18$  simulated cases available in the test case library \cite{websiteForcedOsi}. As presented in Table \ref{tab:SINDyRPCAcomparison},
        both the proposed algorithm and the RPCA method are successful in most cases. However, the proposed algorithm fails in Case $F7-1$, while the RPCA method fails in Case $F7-1$, Case $FM6-2$ (Case B), and Case $FM7-1$ (Case C).
        
        In Case $F7-1$, two sinusoidal forced oscillations are injected into the excitation systems of the generator at bus 79 and that at bus 118, respectively, to cause resonances with two  different modes at frequencies $0.65$ Hz and $0.43$ Hz. Particularly, the oscillation magnitude of the $0.65$ Hz forced signal injected to the excitation system of  the generator at 79 is only $0.006$ p.u, a very small magnitude. As shown in Fig.\,\ref{fig:F7_1_SINDyRPCA_result}, the proposed algorithm fails to capture the oscillation of $0.65$ Hz and its source at bus 79, while captures the oscillation of $0.43$ Hz and its source, though both the generator at bus 116 and that at bus 118 are picked. %along with a false source, i.e., bus 118 and 116. \color{black} 
        On the other hand, the RPCA method fails to capture the source of the forced oscillation at $0.43$ Hz while correctly pinpoints the source of the oscillation at $0.65$ Hz, i.e., the generator at bus 79. 
        
        \color{black} Based on the results in Table \ref{tab:SINDyRPCAcomparison}, \color{black}
        another observation found through the analysis is that compared to the RPCA method, the proposed algorithm may perform better in cases where the forced oscillations come from the governor compared to the RPCA method. %than in those where the forced oscillations stem from exciters. 
        The reason is that the proposed algorithm is based on the swing equations in which the forced signal is assumed to affect the mechanical power (see (\ref{eq:swingEquation_1})) that is typically controlled by the turbine governor. Yet, the RPCA method does not assume any physical model and seems to perform less robust in these cases (e.g., $FM$ cases in Table \ref{tab:SINDyRPCAcomparison}). %However, the RPCA method may perform better in cases where oscillation sources are not generators. Further study and comparison are needed. \color{black}
        \begin{table*}[!ht]
         \caption{Single location cases with resonance between forced oscillation and system modes from \cite{realCasePaper} }
        \label{tab:SINDyRPCAcomparison}
        \centering
        \begin{tabularx}{\textwidth}{c c c c c c c}%{@{}l*{5}{C}c@{}}
        \toprule
        \multicolumn{3}{|@{}c|}{ \color{black} Case prefix "F" and "FM" is for FO in the excitation and governor system\color{black}} & \multicolumn{2}{@{}c|}{SINDy Result}  & \multicolumn{1}{@{}c|}{\qquad \quad RPCA Result \qquad\quad} \\ 
        \midrule
        Cases Name  &  Description  & Source bus  &  Est Source bus  & Est Freq  &  Est Source bus \\ \midrule
        F1 &  Resonance with local 0.86Hz mode &  4  & 4   &  0.86   & 4  \\ 
        F2 &  Resonance with local 0.86Hz mode &  79  & 79   &  0.86   & 79  \\
        F3 &  Resonance with inter-area 0.37Hz mode &  77  & 77   &  0.36   & 77  \\ 
        F4-1 & Below natural 0.84Hz mode &  79  & 79   &  0.8 & 79  \\ 
        F4-2 & Between natural 0.84Hz and 0.86Hz modes &  79  & 79   & 0.86 & 79  \\ 
        F4-3 & Higher than natural 0.86Hz mode &  79  & 79   & 0.9 & 79  \\ 
        F5-1 & Below natural 0.44Hz inter-area mode &  79  & 79   &  0.42 & 79  \\ 
        F5-2 & Between natural 0.44Hz and 0.47Hz inter-area modes&  79  & 79   &  0.46 & 79  \\ 
        F5-3 & Higher than natural 0.47Hz inter-area mode &  79  & 79   &  0.5 & 79  \\ 
        F6-1 & Rectangular wave of 0.1Hz &  79  & 79   & 0.3 & 79  \\ 
        F6-2 & Rectangular wave of 0.2Hz &  79  & 79   & 0.2 & 79  \\ 
        F6-3 & Rectangular wave of 0.4Hz &  79  & 79   & 0.8 & 79  \\ 
        %\rowcolor{gray!20}
        F7-1 & \makecell{Two resonance modes 0.65 Hz (bus 79) and  \qquad\qquad\quad\\ 0.43 Hz (bus 118)}  &  79,118 & \cellcolor{gray!20} 116,118  & \cellcolor{gray!20} both at 0.43 & \cellcolor{gray!20} 79  \\ 
        F7-2 &  Two signals resonance with same mode 0.43 Hz &  70,118  & 70,118  & both at  0.43 & \cellcolor{gray!20} 65  \\ 
        FM1 &  Resonance with local 0.86Hz mode &  4   & 4   &  0.86   & 4  \\ 
        FM3 &  Resonance with inter-area 0.37Hz mode &  77  & 77   &  0.37   & 77  \\ 
        %\rowcolor{gray!20}
        FM6-2 & Rectangular wave of 0.2Hz &  79  &   79 & 0.2 & \cellcolor{gray!20} 35   \\ 
        %\rowcolor{gray!20}
        FM7-1 & \makecell{Two resonance modes 0.65 Hz (bus 79) and  \qquad\qquad\quad\\ 0.43 Hz (bus 118)}  &  79,118  &  79,118 & \makecell{0.657 Hz (bus 79) \\ 0.43 Hz (bus 118)}  & \cellcolor{gray!20} 79  \\ 
        \bottomrule
        \end{tabularx}
        \end{table*}
        %\color{red} will Table I includes all cases? 
        \color{black}

        %%%%%%%%%%%%%%%%%%%%%%%%%%%%%%%%%%%%%%%%%%%%%%%%%%%%%%%%%%%%%%%%%%%%%%%%%%%%%%%%%%%%%%%
        \color{black}
        \subsection{Case D: Forced Oscillation with  a Sudden Frequency Change} %in IEEE 68 bus system } 
\begin{figure}[!ht]
                \centering
                {\includegraphics[width= 3.6 in] 
                {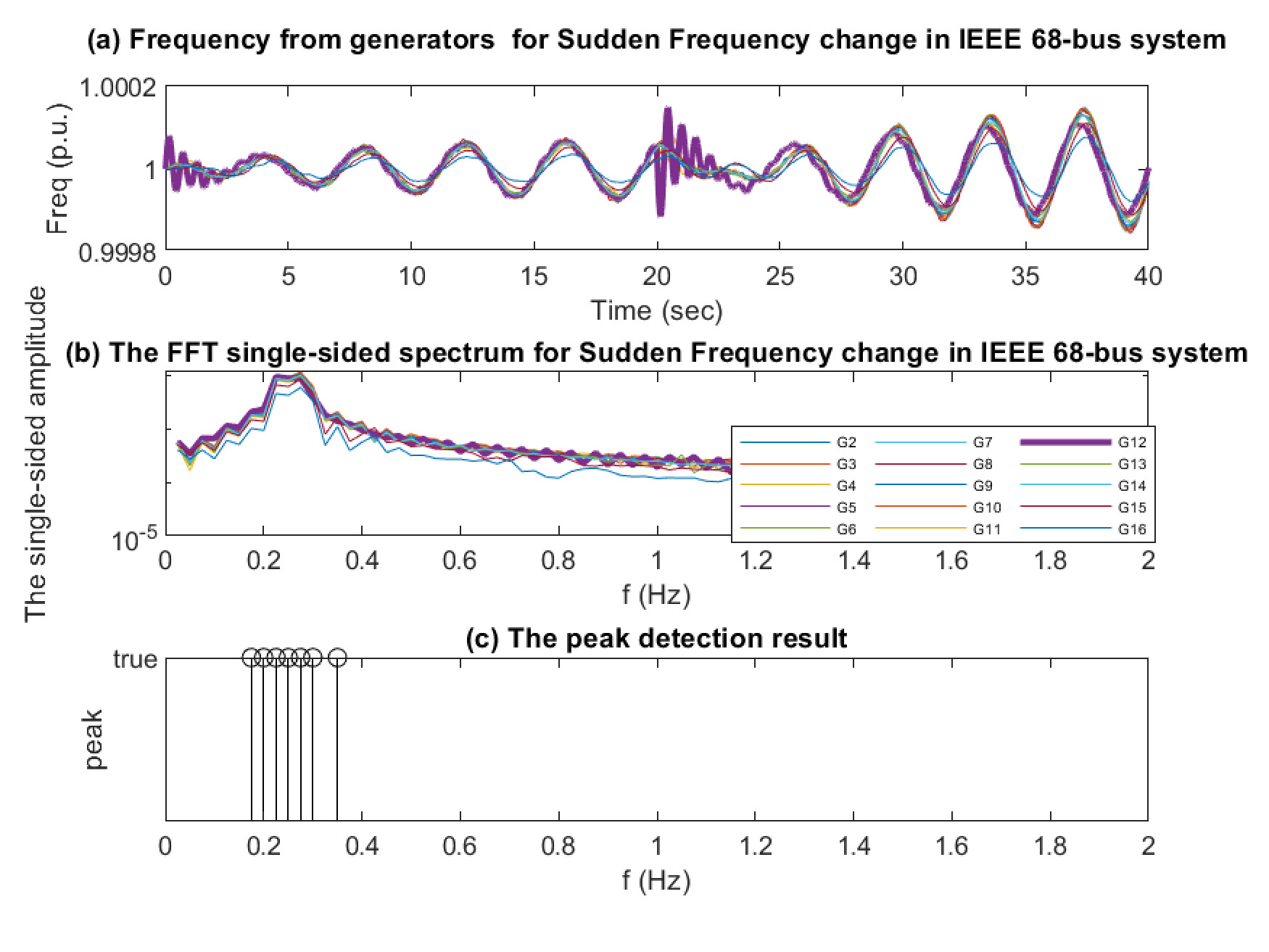}}
                \caption{
                (a). The trajectories of rotor frequencies;  (b) The single-sided spectrum from FFT ; \color{black} (c) The peak frequencies detected for Case D (forced oscillation with a sudden frequency change at 20s).} %in generator $12$.
                %{\color{violet} (Moved the title up)} \color{red} (a) is too close to 1.002\color{black}}
                \label{fig:ChangingAmplitudeFreq_G12_traj}
            \end{figure}

        \vspace{-0.05 in}
        \label{section:frequencieMagnitudeChange}
        To test the performance of the proposed method when the forcing frequency suddenly changes, we 
        conducted some simulations in the IEEE 68-bus system using Power System Analysis Toolbox (PSAT) simulation software \cite{PSATman} as the library\cite{websiteForcedOsi} does not contain such a case. 
        In the IEEE 68-bus system, 16 generators are modelled by the fourth-order model equipped with the standard IEEE type 1 automatic voltage control regulators (AVRs). Generator 1 is selected as the reference generator. The standard deviation of load fluctuation is set to 1 ($\sigma_{load_{i}}$ in (\ref{eq:load})). The proposed algorithm uses the first 40s emulated PMU measurements of rotor angles and frequencies of Generator 2-16 with a sampling rate of $60$ Hz %obtained from PMU measurements of $60$ Hz 
        to build the measurement matrix $X$ and $\dot X$. 
        At the beginning, the forced oscillation with a frequency of $0.24$ Hz and a peak strength of $5\%$ of the mechanical power input is injected into Generator $12$. Later at 20s, the forced oscillation suddenly switches to $0.27$ Hz with a peak strength of $10\%$ of the mechanical power input. The trajectories of rotor frequencies is presented in Fig.\,\ref{fig:ChangingAmplitudeFreq_G12_traj} (a). 
        %The measurement matrix $X$ is built using the data from the first 40 seconds of the rotor angle and frequency data.
        %\color{violet}
        The single-sided amplitude spectrum for the trajectories is shown in Fig.\,\ref{fig:ChangingAmplitudeFreq_G12_traj} (b).
        After removing $63$ duplication, $6$ candidate frequencies are collected from peak detection process in $\Omega_F$ including $0.25$ Hz and $0.275$ Hz that are close to the true forcing frequencies %as $[0.25$ $0.025$ $0.275$ $0.225$  $0.05$  $0.3]$ Hz which is 
        as shown in  Fig.\,\ref{fig:ChangingAmplitudeFreq_G12_traj} (c). 
        \color{black}
        %\color{red}add spectrum, adjust explanation\color{blue}
        
        It is determined by the proposed algorithm that there is forced oscillation from Generator $12$ at $0.25$ Hz as shown in Fig.\,\ref{fig:ChangingAmplitudeFreq_G12_SINDyResult2} (right).
        %The peak selection process selects 6 forced oscillation candidates.
        %The exact candidate frequencies captured are following: $[0.25$ $0.025$ $0.275$ $0.225$  $0.05$  $0.3]$ Hz. 
        However, since there is a sudden change of forcing frequency, %\color{violet} 
        the calculated $\zeta$ matrix %{\color{violet} (I think it should be $\zeta$, but it is calculated from  $\Xi_{ab}$)}{\color{red} $\Xi_{ab}$?} 
        has peaks at both $0.25$ Hz and $0.275$ Hz %is not sparse for the frequency row index in source bus generator $12$ column 
        as shown in Fig.\,\ref{fig:ChangingAmplitudeFreq_G12_SINDyResult2} (left), indicating a potential change of frequency in the time window applied. Nevertheless, the source Generator 12 is located correctly by the proposed method.  %The second highest peak is $0.275$ Hz. (R2-6)  \color{red} I don't see it is less sparse compared to figures in the previous cases? 
        \color{black} 
            \begin{figure}[!ht]
                \centering
                {\includegraphics[width= 3.6 in]{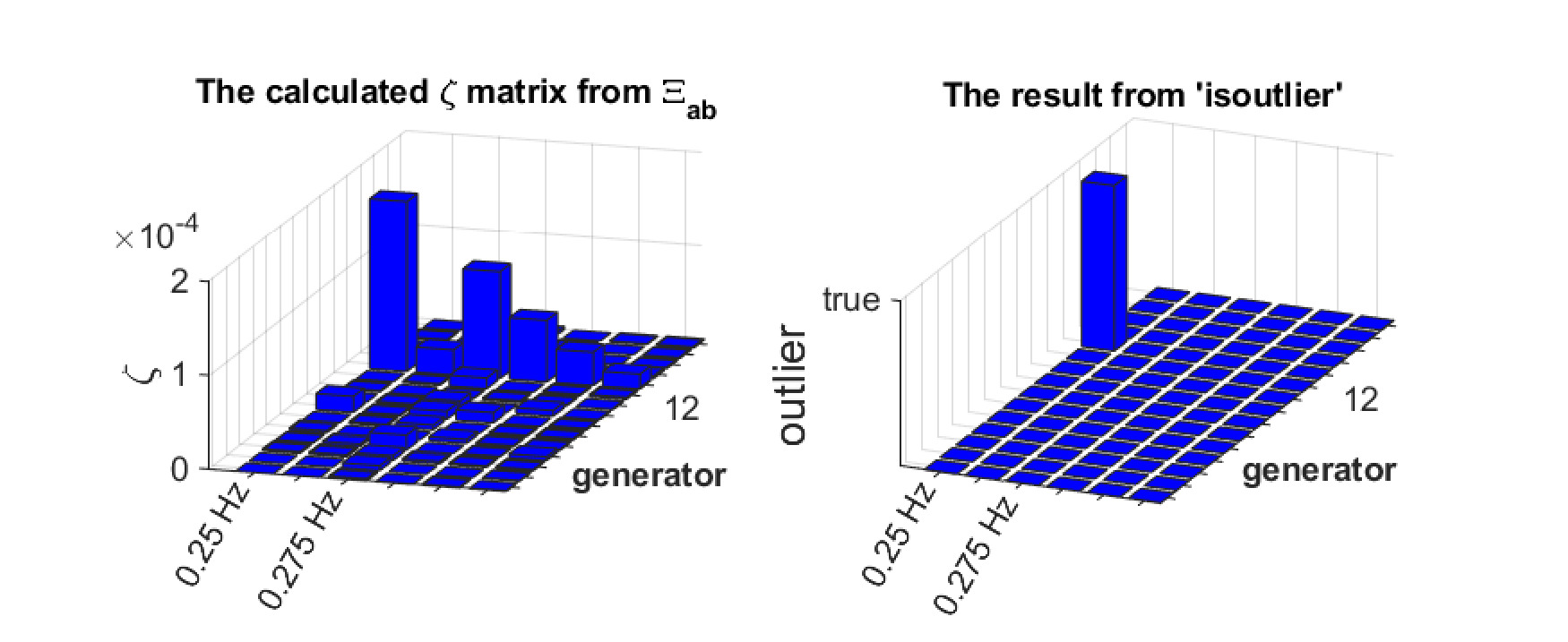}}
                \caption{$\zeta$ (left) and outlier (right) result for Case D (forced oscillation with a sudden frequency change at 20s).} \vspace{-0.15in}
                \label{fig:ChangingAmplitudeFreq_G12_SINDyResult2}
            \end{figure}
        
        %\color{blue}
        %\color{red} TBD\color{blue}In addition, the performance of the proposed algorithm for locating forced oscillations has been tested in $45$ different scenarios where the oscillation frequency is exact at system mode frequency as $0.243$ Hz, near system mode as $0.25$ Hz, or far away from the system mode as $0.5$ Hz in all $15$ generators.
        %The results are correct for all cases. (R3)
        %\color{black}

        %%%%%%%%%%%%%%%%%%%%%%%%%%%%%%%%%%%%%%%%%%%%%%%%%%%%%%%%%%%%%%%%%%%%%%%%%%%%%%%%%%%%%%%
        \color{black}
        \vspace{-0.13 in}
        \subsection{Case E: Under Load Power Factor Variation} %in IEEE 68 bus system } 
        \label{section:powerFactorChange}
        
        Although constant power factor is assumed in (\ref{eq:load}), it may not be true in practical applications. To test the performance of the proposed method under load power factor variation, some simulations in the IEEE 68-bus system were carried out in PSAT. %with dynamic load model using Power System Analysis Toolbox (PSAT) simulation software \cite{PSATman}.
        %as the library\cite{websiteForcedOsi} does not specify the load condition. 
        The generators and their control set up are the same as in Section \ref{section:frequencieMagnitudeChange}. 
        In addition, all $34$ loads in the system have been replaced by voltage dependent loads. The active and reactive power of the dynamic loads are set as random variables  following normal distributions with standard deviations equal to $0.01\%$ %(the border value between stable and unstable cases is $0.03\%$.) 
        of their original power flow solutions.
        %An assertion for $0.95$ power factor check is also added in the simulation.
        If the random load changes cause the power factor falls below $0.95$ any time in the simulation, the reactive power will be adjusted accordingly to maintain $0.95$ power factor, mimicking the power factor correction action.
        
\begin{figure}[!ht]
                \centering
                {\includegraphics[width= 3.6 in] %height= 1.7 in]
                {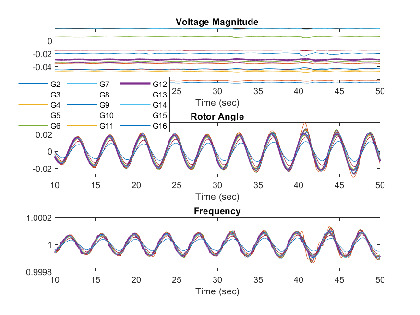}}
                \caption{
                The trajectories of voltage magnitudes (top), rotor angles (mid), and rotor frequencies (bottom) for Case E (under load power factor variations).} \vspace{-0.2in}%in generator $12$.
                %{\color{violet} (Moved the title up)} \color{red} (a) is too close to 1.002\color{black}}
                \label{fig:NO_G12_PowerFactorChange_traj}
            \end{figure}

        Similar to Section \ref{section:frequencieMagnitudeChange}, we assume that the forced oscillation of $0.25$ Hz (near the system mode $0.243$ Hz) is injected into the system through the mechanical power of Generator 12 with a peak strength of $5\%$ of the mechanical power output. %due to, for example, the malfunctions of generator governor control. 
        %The proposed algorithm uses $40$s, $60$Hz emulated rotor angles and frequencies of Generator 2-16 obtained from PMU measurements to build the measurement matrix $X$ and $\dot X$.
        %Particularly, the finite difference method is applied to estimate $\Delta \dot{\omega}_i(t)$, i.e., part of $\dot X$. 
        %The results for one case in which the forced oscillation frequency is $0.25$ Hz near the system mode  $0.243$ Hz are shown here. 
        %The forced oscillation is injected into the system from Generator $12$ ($G12$) with a peak strength of $5\%$ of generator mechanical power output.
         Fig.\, \ref{fig:NO_G12_PowerFactorChange_traj} (top) shows that the voltage magnitudes are varying randomly due to the variations of both active and reactive power (i.e., power factors). %random power factor changes. 
        The trajectories of rotor angles and frequencies are presented in  Fig.\, \ref{fig:NO_G12_PowerFactorChange_traj} (mid) and (bottom).
        The proposed algorithm uses $40$s, $60$ Hz emulated rotor angles and frequencies of Generator 2-16  to build the measurement matrix $X$ and $\dot X$. Likewise, the finite difference method is applied to estimate $\Delta \dot{\omega}_i(t)$. It can be seen from Fig.\, \ref{fig:ChangingAmplitudeFreq_G12_SINDyResult2} (right) that the proposed method can determine accurately that the forced oscillation is from Generator $12$.  %as shown in  . %though the identified frequency is $0.25$ Hz,
        The results align well with the simulation setup.
        From the simulation results, we observe that %small power factor variations have limited impact on the rotor dynamics as shown in  Fig.\, \ref{fig:NO_G12_PowerFactorChange_traj} (mid) and (bottom) and 
        the proposed algorithm is capable of working under small power factor variations.\color{black}
        \color{black} 
            \begin{figure}[!ht]
                \centering
                {\includegraphics[width= 3.6 in]{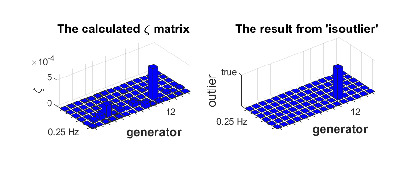}}
                \caption{$\zeta$ (left) and outlier (right) result for Case E (under load power factor variations).} \vspace{-0.2in}
                \label{fig:ChangingAmplitudeFreq_G12_SINDyResult2}
            \end{figure}

        %%%%%%%%%%%%%%%%%%%%%%%%%%%%%%%%%%%%%%%%%%%%%%%%%%%%%%%%%%%%%%%%%%%%%%%%%%%%%%%%%%%%%%%
        \color{black}
        %\vspace{-0.1 in}
        \subsection{Case F: Testing with sixth-order generator models in the WECC 240-bus system} %in IEEE 68 bus system } 
        
        For further validation, we carried out simulations in the WECC 240-bus system in TSAT software \cite{websiteTSAT}, which was developed by National Renewable Energy Laboratory (NREL) \cite{yuan2020developing}.  
        %In order to test the proposed algorithm under an extensive system with a higher-order generator model, a set of TSAT simulations is carried out with a WECC 240-bus test system case which National Renewable Energy Laboratory (NREL) develops based on reference \cite{yuan2020developing}. 
        %\color{violet} 
        The system has $243$ buses, $146$ generating units at $56$ power plants.
        All synchronous generators are modelled as the sixth-order GENROU synchronous generator model.
        The small-signal analysis result can be found \cite{websiteContest}.
        The forced oscillations with the peak strength of $40$MW are injected into the system through the TGOV1 governor models or SEXS excitation models.  %output as the malfunctions of gas or hydro generator control.
        % with SEXS excitation system model and TGOV1 governor models. 
        %\color{red} are all generators modeled as 6th-order model? %\color{violet} (Yes, all synchronous  generators are modelled  as GENROU. but only gas and hydro are using TGOV1 governor, In addition, GENROU is 6th order model as \cite{genrou6th}) 
        The forced oscillation frequencies are set at the exact mode frequencies, aiming to create resonance cases. Three system mode frequencies at $1.1819$ Hz, $0.803$ Hz and $0.5774$ Hz are picked, the damping ratios of which  
        %with the near system mode pair frequencies as $1.2$ Hz, $0.825$ Hz, and $0.55$ Hz.
        %The damping ratio for system modes 
        are $4.5\%$, $6.5\%$, and $8.1\%$, respectively. 
        %The largest oscillation amplitudes appears at non-source buses for all cases, except for the cases involving %the forced signal is injected to 
        %bus $2634$ and $7031$. \color{blue}
        %The small-signal analysis result can be found in \cite{websiteContest}.
        %$8$ generators (as bus $1034$, $1431$, $2634$, $4132$, $5032$, $6433$, $6533$, and $7032$) are randomly selected from different areas for the injection.
        Additional cases with the rectangular forced signal with a fundamental frequency of $0.37$Hz are also tested.
        %Some significant harmonic elements are $0.74$ Hz, $1.11$ Hz, and $1.48$ Hz.
        %When the rectangular forced signal is injected to bus $6335$, the second harmonic $0.74$ Hz resonates with the system mode and produces the largest oscillation amplitude at a non-source bus.\color{red} which case?\color{blue} %in the power flow trajectory, which is not at the source.
        $11$ generators are randomly selected from different areas for the injection of forced oscillations. 
        The largest oscillation amplitudes appear at  non-source buses in all cases except the cases %forced signal is injected to bus 
        involving buses $2634$, $6335$, or $7031$. %\color{black} \color{red} which case?\color{blue}
        The proposed algorithm successfully locates the source bus for all cases. 
        The summary of the test results is shown in Table \ref{tab:result_WECC240_2}.
        
        \color{black}
        
        \iffalse
            \begin{threeparttable}[!ht]
                \caption{Result summary for testings in WECC 240-bus}
                \label{tab:result_WECC240}
                \begin{tabular}{c c c c}
                \toprule
                \multicolumn{1}{|@{}c|}{ \quad Source Bus } & 
                \multicolumn{1}{|@{}c|}{  Simulation 1 } & 
                \multicolumn{1}{|@{}c|}{ Simulation 2 } 
                \\ 
                \midrule
                1034 & 0.803 Hz, 0.825 Hz & 0.5774 Hz, 0.55 Hz\\
                1431 & 1.1819 Hz , 1.2 Hz & 0.803 Hz,  0.825 Hz\\
                2634 & 0.803 Hz,  0.825 Hz & 1.1819 Hz,  1.2 Hz\\
                4132 & 1.1819 Hz , 1.2 Hz & 0.803 Hz,  0.825 Hz\\
                5032 & 0.803 Hz,  0.825 Hz &  0.5774 Hz, 0.55 Hz\\
                6433 & 0.5774 Hz,  0.55 Hz & 1.1819 Hz,  1.2 Hz\\
                6533 & 0.803 Hz,  0.825 Hz & 0.5774 Hz,  0.55 Hz\\
                7032 & 0.5774 Hz , 0.55 Hz & 0.803 Hz,  0.825 Hz\\
                \bottomrule
                \end{tabular}
            \end{threeparttable}
        \color{red} also present a figure for one particular case as shown in Fig. 14. \color{black}
        \fi
        
        \small
            \begin{threeparttable}[!ht]
                \caption{Result summary for tests in WECC 240-bus}
                \label{tab:result_WECC240_2}
                \begin{tabular}{c c c c }%{@{}l*{5}{C}c@{}}
                \toprule
                \multicolumn{2}{|c|}{Simulation setup} & \multicolumn{2}{c|}{SINDy Result} \\ 
                \midrule
                Source & Description   &  Source&Freq(Hz) \\ \midrule
                1034& Sine wave of 0.5774Hz from exciter   & 1034   &  0.577 \\ 
                1431& Sine wave of 0.803Hz from governor  & 1431   &  0.803 \\
                2634& Sine wave of 1.1819Hz from governor   & 2634   &  1.18 \\ 
                3933& Sine wave of 0.5774Hz from exciter &  3933   &  0.577\\ 
                4132& Sine wave of 0.803Hz from governor  & 4132   & 0.803 \\ 
                5032& Sine wave of 1.1819Hz from exciter &  5032    & 1.18 \\ 
                6533& Sine wave of 0.5774Hz from governor &  6533   &  0.577 \\ 
                7031& Sine wave of 0.803Hz from exciter  & 7031   &  0.803 \\ 
                7032& Sine wave of 1.1819Hz from governor  & 7032   &  1.18\\
                2030& \makecell{Rectangular wave of 0.37Hz \\ from exciter } &  2030   & 0.74  \\ 
                6335& \makecell{Rectangular wave of 0.37Hz \\from governor} &  6335   & 0.37 \\
                \bottomrule
                \end{tabular}
            \end{threeparttable}
            %\color{red} will Table I includes all cases? 
        \normalsize
        \color{black}

%%%%%%%%%%%%%%%%%%%%%%%%%%%%%%%%%%%%%%%%%%%%%%%%%%%%%%%%%%%%      
    \section{\color{black}\uppercase{testing in actual oscillation events}\color{black}}\label{section:real case testing}
        
\begin{figure}[!ht]
                \centering
                \includegraphics[width=3.2 in, height= 2 in]{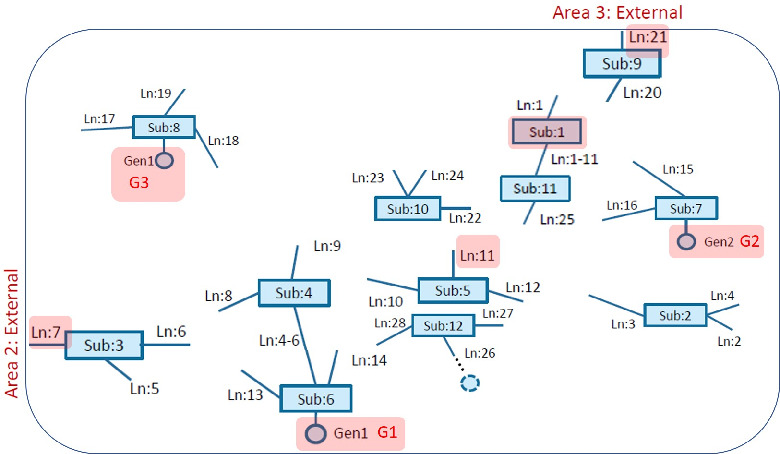}
                \caption{Partial map for ISO New England system from \cite{websiteForcedOsi}.}
                \label{fig:mapForISONE} \vspace{-0.15in}
            \end{figure}

        \color{black}
        In the next three subsections, the performance of the proposed algorithm is tested using the measurements from the real-life forced oscillation events in the system of ISO-NE \cite{websiteForcedOsi}. The partial map for the system is shown in Fig.\,\ref{fig:mapForISONE}. 
        %\color{blue} (18June2021) 
        There are three generators inside the study area: G1 %$1$ is connected with substation $6$ 
        (Gen$1$ at Substation 6), G$2$ (Gen$2$ at Substation 7), and G$3$ (Gen $1$ at Substation 8). %is connected with substation $7$, and another Generator $1$ (G$1$S8) which is connected with substation $8$.
        The study area is connected to two external areas \say{Area 2} and \say{Area 3}. 
        In addition, the boundary buses---Line $7$ at Substation $3$ (Ln:$7$) and Line $21$ at Substation $9$ (Ln:$21$)---are used to represent aggregations of external areas. % used %as the generator bus for 
        %used by the algorithm to represent aggregations of external areas. %handle the forced oscillation from external areas. 
        As rotor angles are  unavailable, the measurements of voltage angles and frequencies of all generator substation terminal buses and the aforementioned two boundary buses are used. Such change may affect the estimation of the Jacobian matrix, but the coefficient matrix $\Xi_{ab}$ still preserves the information of forced oscillation. 
        %\color{blue} (18June2021) 
        The bus---Line $11$ at Substation $5$ (Ln:11) is chosen as the reference bus.
        %\color{black}
        %No de-trending and de-noise methods are applied. 
        The raw PMU data are fed directly to the proposed algorithm.
        
        Moreover, the success of the proposed method using data from actual oscillation events demonstrates its robustness with respect to measurement noise, general drift of measurements, and other imperfections in practical applications. 
        %The fast computational speed indicates a good feasibility of the proposed method in online applications. 
        In the end, we will also discuss the limitations of the proposed algorithm. %drawback from some unusual real event cases.
        \color{black}
        
        %%%%%%%%%%%%%%%%%%%%%%%%%%%%%%%%%%%%%%%%%%%%%%%%%%%%%%%%%%%%%%%%%%%%%%%%%%%%%%%%%%%%%%%%%%%% 
        \vspace{-0.2in}
        %\subsection{Case D: Validation in an Actual Forced Oscillation Event With Strong Forcing Strength} %Performance evaluation with real forced oscillation cases}
        \subsection{\color{black}ISO-NE Case 3 in \cite{websiteForcedOsi}: an Actual Forced Oscillation Event With Strong Forcing Strength\color{black}} %Performance evaluation with real forced oscillation cases}
        \label{section:real case}
        %---real case 1 (regular oscillation) ---%
        %\color{red} please change figure titles for Fig. 14 and onward\color{black}
        
\begin{figure}[!ht]
                \centering
                \includegraphics[width=3.6in]{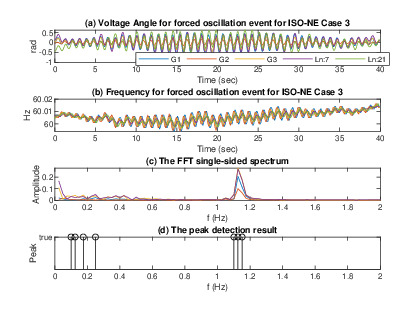}  
                \caption{(a). The trajectories of voltage angles; (b) The trajectories  of  voltage  frequencies;\color{black}(c) The single-sided spectrum from FFT;\color{black}(d) The peak frequencies detected for \color{black} ISO-NE Case 3 in \cite{websiteForcedOsi}\color{black}. } \vspace{-0.2in}%the July 20th, 2017 event. }
                \label{fig:realCase3_DataPeakPlot}
            \end{figure}

        The first forced oscillation event (ISO-NE Case 3 in \cite{websiteForcedOsi}) under study occurred on July 20th, 2017. It was a regional oscillation event. The modes were widely observed within the system. The peak-to-peak magnitude of the oscillation reached 115 MW with a peak load of about 26,000 MW. 
        The oscillation frequency was 1.13 Hz. The source of the oscillation was close to G2. 
        Fig.\,\ref{fig:realCase3_DataPeakPlot} (a)-(b) present $40$s measurements  %window 
        of voltage angles and frequencies of all generators' terminal buses fed to the proposed algorithm. %as shown in  Fig.\,\ref{fig:realCase3_DataPeakPlot} (a)-(b).
        The single-sided spectrum of all measurements obtained from the FFT %all generators 
        are presented in Fig.\,\ref{fig:realCase3_DataPeakPlot} (c). Fig.\,\ref{fig:realCase3_DataPeakPlot} (d) shows the 7 candidate frequencies identified. %\color{red} why 7? I don't see peaks around 0.2 Hz at all \color{blue} (There is a peak in 0.2 Hz, but it is really small.) \color{black}
            \begin{figure}[!ht]
                \centering
                \includegraphics[width=3.6in]{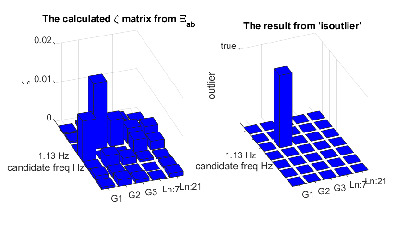}
                \caption{$\zeta$ (left) and outlier (right) result for  \color{black} ISO-NE Case 3 in \cite{websiteForcedOsi}\color{black}.} %\vspace{-0.2in} %the July 20th 2017 event.} 
                \label{fig:realCase3_ResultPlot}
                %\vspace{-2in}
            \end{figure}
            
        Next, the coefficient matrix $\Xi$ is estimated.   
        The indexes $\{\zeta\}$ are calculated and the outliers are identified as presented in Fig.\,\ref{fig:realCase3_ResultPlot}, in which a spike appears. It indicates that there  %A spike appears, indicating there 
        is a forced oscillation at $1.13$ Hz from G$2$. The results well align with the identified cause of this event mentioned above, showing that the proposed algorithm can successfully locate the source in practical applications despite measurement noise, general drift of the data, etc. Furthermore, the algorithm takes only $1.77$s for the whole process. In contrast, the RPCA method identifies the bus---Line $21$ at Substation 9 (Ln:$21$) as the source, which is unfortunately incorrect. %might in the external Area$3.\color{black}
        
        %%%%%%%%%%%%%%%%%%%%%%%%%%%%%%%%%%%%%%%%%%%%%%%%%%%%%%%%%%%%%%%%%%%%%%%%%%%%%%%%%%%%%%%%%%%%%%%
        \vspace{-0.15in}
        %\subsection{Case E: Validation in an Actual Forced Oscillation Event With Low Forcing Strength}
        \subsection{\color{black}ISO-NE Case 5 in \cite{websiteForcedOsi}: an Actual Forced Oscillation Event With Low Forcing Strength\color{black}}
            
        %---real case 2 (weak oscillation) ---%
        Another case we studied was the forced oscillation event in  the system of ISO-NE on January 29th, 2018 (ISO-NE Case 5 in \cite{websiteForcedOsi}). Again, the oscillation was caused by G2. However, the peak-to-peak magnitude was 15 MW, much smaller than that in the previous case. The forced oscillation frequency was around 1.587 Hz \color{black} with a potential graduate frequency variation between 1.57 Hz and 1.63 Hz.\color{black} %However, it also contains a gradual frequency change between 1.57 Hz and 1.63 Hz. \color{black} %\color{black} %1.57 Hz. 
        
\begin{figure}[!ht]
                \centering
                \includegraphics[width=3.6in]{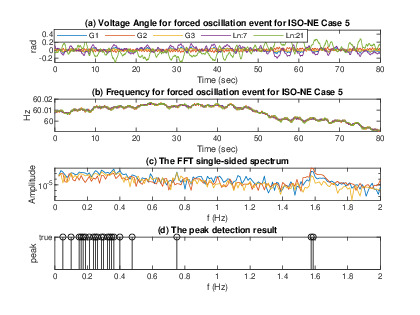}
                \caption{(a). The trajectories of voltage angle; (b) The trajectories  of  voltage  frequencies;\color{black}(c) The single-sided spectrum from FFT;\color{black} (d) The peak frequencies detected for  \color{black} ISO-NE Case 5 in \cite{websiteForcedOsi}\color{black}.} %the January 29th 2018 event.}
                \label{fig:realCase5_DataPeakPlot}
            \end{figure}

        As the oscillation magnitude is small, %To apply the proposed algorithm, 
        measurements of $80$s rather than the default $40$s used in previous cases of voltage angles and frequencies of all generators' terminal buses are used to capture the forced oscillation frequency candidates. %\color{red} 23 seems to be too much \color{blue} (Signals in the channle is messy, do you mean I need to change the peak detection method) \color{black}
        %\color{blue} (11Jane2021) instead of the default $40$s measurements for the previous cases to pinpoint the forced oscillation frequencies. \color{red} how many measurements are used in the previous case?\color{black} %improve the accuracy of frequency identification. %to build the measurement matrix $X$ and $\dot X$. 
        As shown in Fig.\,\ref{fig:realCase5_DataPeakPlot} (a)-(b), the forced oscillations can be hardly observed in the time-domain plots.  Fig.\,\ref{fig:realCase5_DataPeakPlot} (c) presents the single-sided spectrum obtained from the FFT. %in which no significant peaks can be identified.  %from Fig.\,\ref{fig:realCase5_DataPeakPlot} (c). 
        With the help of the z-score peak detection method, 23 candidate frequencies are identified as shown in Fig.\,\ref{fig:realCase5_DataPeakPlot} (d).
            \begin{figure}[!ht]
                \centering
                \includegraphics[width=3.6in]{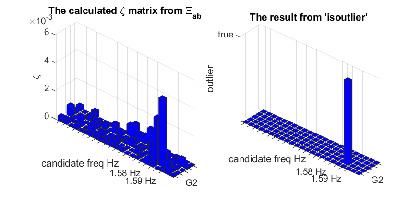}
                \caption{$\zeta$ (left) and outlier (right) result for  \color{black}ISO-NE Case 5 in \cite{websiteForcedOsi}\color{black}. } \vspace{-0.2in}%the January 29th 2018 event.}
                \label{fig:realCase5_ResultPlot}
            \end{figure}  
            
        %Next, the coefficient matrix $\Xi$ is estimated. Particularly, $\|\Xi_{ab}\|_2=0.00244$ that still stands out. 
        \vspace{0.07in}
        The indexes $\{\zeta\}$  are further calculated as well as the outliers, as presented in Fig.\,\ref{fig:realCase5_ResultPlot}.  %\color{red} Fig wrong. please double check all figure label ref \color{black}. 
        The results imply that there is a forced oscillation at $1.59$ Hz from G2, well consistent with the identified cause. 
        Moreover, the algorithm takes only $2.08$s for the whole process. 
        
        %From the above two case studies in actual oscillation events, it can be observed that the proposed method can accurately and efficiently %diagnose oscillation mechanism and 
        %locate forced oscillation sources using a short window (40-80s with 30 Hz sampling frequency) of raw PMU data in practical applications. %without pre-processing or frequent tuning. 
        %The computational time is only around $3$s. %The small sample size and fast computational speed  demonstrate the feasibility and efficacy of the proposed method in practical online applications. 

        %%%%%%%%%%%%%%%%%%%%%%%%%%%%%%%%%%%%%%%%%%%%%%%%%%%%%%%%%%%%%%%%%%%%%%%%%%%%%%%%%%%%%%%
        %\color{blue}
        \vspace{-0.15in}
        \subsection{Further Validations of the Proposed Method in Actual Forced Oscillation Events} \label{section:FurtherValidationsReal}
        %Performance in Cases Where the Forced Oscillations are Outside the Study Area, the Missing Measurements, and using the Measurements from Transmission Line}
        
        In this section, the other cases (Case $1$, $2$, $4$ and $6$ in ISO-NE) available in the library \cite{websiteForcedOsi} are tested, in which the source is outside the study area or there are missing measurements. 
        %In addition to the forced oscillation events occurred in the study area. The proposed algorithm is also tested for the cases which possess the forced oscillation passing by other area from the library  \cite{websiteForcedOsi} as ISO-NE Case $1$, $2$, and $4$. 
        
        In Case $1$ and $4$, the forced oscillation source is in the external Area $2$. Case $1$ occurred on June 17th, 2016, with a peak-to-peak magnitude up to $27$ MW. Case $4$ occurred on Feb. 14th, 2018, with a peak-to-peak magnitude up to $10$ MW. 
        %\color{blue}  
        Both the proposed algorithm and the RPCA method identify that the border bus of Area $2$---Line $7$ at Substation 3 (Ln:$7$) is the source. \color{black}
        
        %\color{blue} (11Jane2021)
        In Case $2$, the data from G$1$ %$1$S$6$ and 
        and G$3$ %$G$1$S$8$ 
        is missing. In fact, all measurements of Substation 6 and Substation 8 are unavailable. Case $2$ occurred on October 3rd, 2017, with a peak-to-peak magnitude up to $130$ MW. To apply the proposed algorithm, the data from Line $4$-$6$ at Substation $4$  that is a direct connection to Substation $6$ is used to replace $G1$, %G$1$S$6$, 
        while the channel for G$3$ has to be dropped. %\color{red}Whether any measurements of Sub 8 are available? \color{black} \color{red} relation to G3? closest? is there a typo in the figure \color{black} 
        The proposed algorithm declares Line $21$ at Substation 9 as the source, indicating that the forced oscillation may be from Area 3, which is consistent with the identified cause. The RPCA method also correctly pinpoints the source in this case.
        
        The results of the proposed algorithm and the RPCA method for all actual oscillatory events from \cite{websiteForcedOsi} are summarized in Table \ref{tab:RealCases}. It is observed that SINDy is able to accurately identify the sources in most cases except Case 6 in which the forced oscillations can be hardly observed in any voltage angle measurements. %It is observed that the proposed method fails in Case 6 when the forced oscillation is hardly observed in voltage angle measurements due to the physical locations of the measurement units are close. 
        %\color{blue} (18June2021)
        Case $6$ occurred on \color{black} June 20th, \color{black} 2019, with peak to peak magnitude up to $9$ MW. The PMU measurements are obtained from $4$ transmission lines connected to Substation $1$ (though only two lines are marked on the map in Fig. \ref{fig:mapForISONE}\cite{websiteForcedOsi}). %\color{red} There is no substation 1 in map 
        \color{black}
        %The oscillations in the voltage angles are holding constant with some small changes as a jagged lines with peak to peak difference is less than $0.001$ rad.
        Unfortunately, the proposed algorithm cannot establish a relationship between the voltage angles and bus frequencies, leading to a regularized result (all elements are non-zero and equally high) in $\Xi_{ab}$.
        %\color{blue}
        Therefore, the source is not located. This indicates that it might be insufficient to use classic swing equation to describe the voltage angle and frequency relationship for a bus not directly connected to a generator, especially when the generators are far away from the bus or when the forced oscillation magnitude is small. %due to the close distance between sensor unit.
        \color{black} Or more advanced signal processing techniques are needed to remove all the environmental noise and reappear the forced oscillation signals in the trajectories.
        \color{black}%for the data from the no generator connected bus. 
        Besides, further efforts need to be devoted to searching for more appropriate nonlinear relationships used by SINDy when the measurements are far away from generators or when the forced oscillation sources are not generators (e.g., HVDC control). \color{black} %and building more terms to enrich the SINDy library can be developed in future.
        %\color{red} please update \color{black}

        %All estimated forced oscillation frequencies from above cases are a bit off from the real forced frequencies.
        %Especially, the estimated oscillation frequency in Case $4$ is ten times smaller than the real oscillation frequency. 
        %Such error is related to the frequency resolution of the FFT process. 
        %The resolution for spectrum of FFT has a positive correlation to the sampling frequency of the data matrix, and a negative correlation to the length of the data matrix.
        %A low sampling frequency with a short time data matrix causes a relatively large spectrum resolution. 
        %This might lead to the exact oscillation frequency is missing in the spectrum grid. 
        %However, the harmonic of the exact oscillation frequency might lay perfectly align with the spectrum grid.
        %Case $4$ suggested that the correct oscillation frequency is obtained by explore harmonic of the algorithm result.

        \begin{table*}[!ht]
         \caption{Cases of actual oscillatory events from \cite{websiteForcedOsi} }
        \label{tab:RealCases}
        \begin{tabular}{c c c c c c}
        \toprule
        \multicolumn{3}{|@{}c|}{Real events} & \multicolumn{2}{@{}c|}{SINDy Result}  & \multicolumn{1}{@{}c|}{\quad RPCA Result |} \\ 
        \midrule
        ISO-NE & Description & Real Location & Est Location & Est Freq (Hz) & Est Location \\ \midrule
        %\rowcolor{gray!20}
        Case 1 &  Near-resonance conditions with natural oscillatory mode 0.27 Hz & Ln:7 (Area 2) &  Ln:7  & 0.275, 0.286   &  Ln:7  \\ 
        Case 2  &  Growing oscillations at 0.08Hz, 0.15Hz and 0.31Hz &  Ln:21 (Area 3)  & Ln:21   &  0.338    & Ln:21  \\
        Case 3  &  Equipment issue in generator 2 has created 1.13  &  G2  & G2  &  1.13  & \cellcolor{gray!20}Ln:21 \\ 
        Case 4  &  Possible resonance case with 0.25Hz mode &  Ln:7 (Area 2)   & Ln:7   &  0.25   & Ln:7  \\ 
        Case 5  &  Local oscillations caused by generator 2  &  G2  & G2   &  1.59   & G2  \\ 
        %\rowcolor{gray!20}
        Case 6  &  PMU measurements from Substation 1  transmission line 1 to 4  &  Substation 1 Ln:2  & \cellcolor{gray!20} Cannot tell   & \cellcolor{gray!20} 0.0107   & Sub1 Ln:2  \\ 
        \bottomrule
        \end{tabular}
        \end{table*}
        \color{black}

%%%%%%%%%%%%%%%%%%%%%%%%%%%%%%%%%%%%%%%%%%%%%%%%%%%%%%%%%%%% 
    \color{black}
    %\vspace{-0.15in}
    \section{comparison with the dissipating energy flow method} %Energy-based Localization Method}
    \color{black}
        The energy flow based forced oscillation locating method is widely used in the utility practice, which is accurate and robust\cite{ISONewEnglandExperience}. In this section, we show that the proposed SINDy algorithm can be a valuable companion to the energy-based method in some tricky cases. 
        %Even if some regions of the typology are missing, the aggregated area contour \cite{DEFContour} can be deployed to help the system operator to locate the source within a small number of bus sets.
        \color{black}
        %As pointed out in \cite{foHuang}, if the source generator is not monitored and erroneous system topology is used, energy flow-based method \cite{6296740} \cite{maslennikov2017dissipating} may report false-positive results. To investigate this, %In contrast, under the same condition, the RPCA method may  proposed SINDy method will  pinpoint the generator closest to the source as the source generator. 
        %we tested the three methods, i.e., the Dissipating Energy Flow method (DEF) in \cite{maslennikov2017dissipating}, the RPCA in \cite{foHuang} and the proposed SINDy for Case $FM1$ in the library. %under some special conditions. 
        %First, We follow the assumption in \cite{foHuang} that PMUs are installed atall generator buses except ones at Buses 4. Buses 7 and 19 are also installed with PMUs. The topology errors occurred between buses 19, 16, and 146.Such topology errors cause DEF-based method to falsely report bus 15 as the source. However, both the proposed algorithm and RPCA-based method suggest the source measurement is at Bus 7 which is the closest bus to the source bus 4. 
        %In addition, 
        Similar to the approach in \cite{foHuang}, 
        %\color{blue}
        %However, we want to show an extreme case that demonstrates the potential advantage of the proposed SINDy algorithm when the measurements of the true source are unavailable.  
        we assume that the measurements of the source generator at bus 4 are not available, while the measurements of all the other generators are available. \color{black}%and no close bus measurement for the source. 
        %The system topology information is also unavailable. %which DEF method can only estimate the dissipate power flow into or out of the generator bus.
        %The closest generators to the source bus 4 are bus 9 and 18, which is shown in the blue circle of the zoom-in WECC 179 map as Fig.\,\ref{fig:Map_FM1_2}. 
        The results from the the Dissipating Energy Flow method (DEF) in \cite{maslennikov2017dissipating}, the proposed SINDy, and the RPCA are presented in Fig.\,\ref{fig:Result_noBus4_FM1_2}, respectively.
        %From the result plots, 
        Both the RPCA and the proposed SINDy algorithm suggest that the sources are  the generators at bus 9 and bus 18 as shown in Fig.\,\ref{fig:Result_noBus4_FM1_2} (bottom left and bottom right). 
        In addition, bus 30 is also among the top three elements in the residual matrix in the RPCA method as shown in Fig.\,\ref{fig:Result_noBus4_FM1_2} (bottom left). In contrast, 
        %Fig.\,\ref{fig:Result_noBus4_FM1_2} (top) shows the flows of dissipating energy over time at the connection points of the power plants to the network.
        the DEF method fails to identify the source as there are no dissipating energy flows from the generator buses to the network (all energy curves %(In the DEF paper, $W_D$ has been called as "dissipating energy flow" or "Net energy flow".)\color{red} make sure if it is power curves or energy curves\color{blue}are 
        near or lower than zero in Fig.\,\ref{fig:Result_noBus4_FM1_2} (top)).  
        The full result summary for the DEF, the RPCA and the proposed SINDy is shown in Table \ref{tab:CompareDEFRPCASINDy},
        \color{black}
        showing that the proposed SINDy algorithm may complement the DEF method when the measurements of
        the true source are unavailable.
        %which demonstrates that the data-driven based forced oscillation locating algorithm can be a good companion to the energy flow based algorithm \cite{ISONewEnglandExperience} and \cite{DEFContour} in a tricky scenario.
        \color{black}

        \color{black}
        
        \begin{table*}[!ht]
             \caption{Result summary for cases $FM1$ with special conditions}
            \label{tab:CompareDEFRPCASINDy}
            \begin{tabular}{c c c c }
            \toprule
            \multicolumn{1}{|@{}c|}{ Case Description } & 
            \multicolumn{1}{|@{}c|}{ \quad DEF Result } &
            \multicolumn{1}{|@{}c|}{  \quad RPCA Result } &
            \multicolumn{1}{|@{}c|}{  \quad SINDy Result }
            \\ 
            \midrule
            \makecell{No measurements from source bus 4, PMUs are installed only in generator bus, 9 and 18 are  the\\  closest generator buses} & 
            Cannot tell &
            Bus 30, 9, and 18 &
            Bus 9,  and 18
            \\ 
            \bottomrule
            \end{tabular}
        \end{table*}

        \iffalse
        \begin{table*}[!ht]
             \caption{Result summary for cases $FM1$ with special conditions}
            \label{tab:CompareDEFRPCASINDy}
            \begin{tabular}{c c c c c c}
            \toprule
            Case Number &
            \multicolumn{1}{|@{}c|}{ Case Description } & 
            \multicolumn{1}{|@{}c|}{  \quad True Source  } &
            \multicolumn{1}{|@{}c|}{ \quad DEF Result } &
            \multicolumn{1}{|@{}c|}{  \quad RPCA Result } &
            \multicolumn{1}{|@{}c|}{  \quad SINDy Result }
            \\ 
            \midrule
            1&
             \makecell{ No measurements from source bus 4, \\
             measurements from close bus 7 is used as replacement. \\  
            There are topology errors between bus 19, 16, and 146.
            } & 
            Bus 7 &
            Bus 15 &
            Bus 7 &
            Bus 7
            \\ \midrule
            2&
            \makecell{No measurements from source bus 4, no replacing measurements. \\PMUs are installed only in generator bus \\ bus 9 and 18 are the closest generator buses} & 
            \makecell{Bus 9, \\ and 18} &
            Cannot tell &
            \makecell{Bus 30, 9, \\ and 18} &
            \makecell{Bus 9, \\ and 18}
            \\ 
            \bottomrule
            \end{tabular}
        \end{table*}
        \fi
            \begin{figure}[!ht]
                \centering
                {\includegraphics[width= 3.5in]{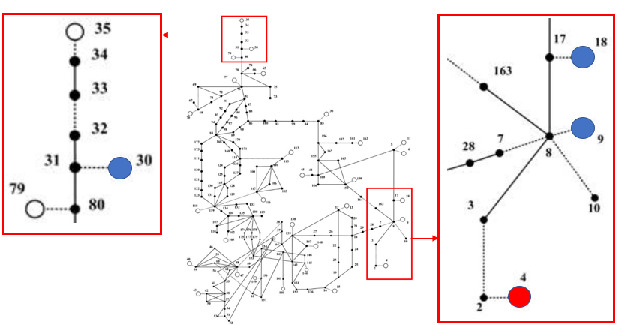}}
                \caption{Map for WECC 179-bus power system used for Case FM1 with no source measurement result \cite{websiteForcedOsi}.}
                \label{fig:Map_FM1_2}
            \end{figure}
            \begin{figure}[!ht]
                \centering
                {\includegraphics[width= 3.6in ]{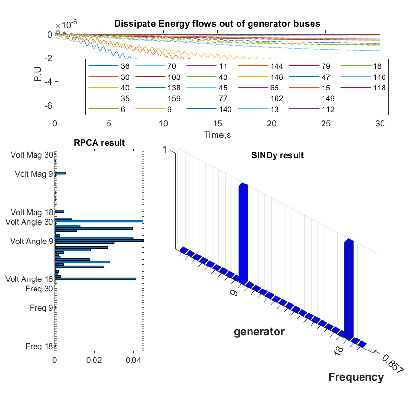}}
                \caption{Result for Dissipating energy flow (DEF) (top), RPCA (bottom left), and SINDy  (bottom right) of all generators except the source bus 4 in case $FM1$ with no source measurement.} \vspace{-0.2in}%\color{red} as least the RPCA and the SINDy method should be presented in the same way as in Fig. 10. Besides, it may be better to use words rather than $W$ and $W_D$ for the DEF\color{black}}
                \label{fig:Result_noBus4_FM1_2} 
            \end{figure}
        \iffalse 
            \begin{figure}[!ht]
                \centering
                \includegraphics[width= 3.5 in, height= 2 in]{figures_Response/FM1_noBus4_DEF_result.jpg}
                \caption{Dissipating energy flow (DEF) of all generators except the source bus 4 in case $FM1$ with no source measurement}
                \label{fig:DEF_noBus4_FM1_2}
            \end{figure}
            \begin{figure}[!ht]
                \centering
                \includegraphics[width= 3.5 in, height= 2 in]{figures_Response/FM1_noBus4_SINDy_result.jpg}
                \caption{$\zeta$ (left) and outlier (right) result for case $FM1$ with no source measurement }
                \label{fig:SINDy_noBus4_FM1_2}
            \end{figure}
            \begin{figure}[!ht]
                \centering
                \includegraphics[width= 3.5 in, height= 2 in]{figures_Response/FM1_noBus4_RPCA_result.jpg}
                \caption{Visualization of the sparse residual matrix $S$ by the RPCA method (left) and the maximum absolute value in each row of $S$ (right) for Case $FM1$ with no source measurement}
                \label{fig:RPCA_noBus4_FM1_2}
            \end{figure}
        \fi

%%%%%%%%%%%%%%%%%%%%%%%%%%%%%%%%%%%%%%%%%%%%%%%%%%%%%%%%%%%%      
    \section{\uppercase{CONCLUSION AND FUTURE WORK}}
    
        In this paper, a purely data-driven algorithm for forced oscillation location has been proposed. Leveraging on the sparse identification for nonlinear dynamical system, the proposed algorithm can extract the dynamics of a power system from PMU data to locate forced oscillation sources.  
        The proposed algorithm is evaluated with good results using the simulated cases and actual oscillation events available in the  IEEE Task Force test cases library \color{black}as well as in the simulated IEEE 68-bus system\color{black}.  %in WECC 179-bus systemIEEE 68-bus system, WECC 179-bus system, and real-world data including some difficult cases such as weak forced oscillation overlapped with poorly damped natural modes. %Despite the good results, some of the topics are still needed for the 
        Future developments such as developing an effective feature library when forced oscillation sources are not generators are needed. %without resorting to FFT are needed. \color{red} Please modify accordingly \color{black}
        
        %\vspace{-0.4 in}

%%%%%%%%%%%%%%%%%%%%%%%%%%%%%%%%%%%%%%%%%%%%%%%%%%%%%%%%%%%%      
    \appendices
    \section{Representing forced oscillations from exciters and turbine governors in swing equations} \label{app:rotorDynamics}
         \color{black} 
        As discussed in \cite{Impedance}, the linearized rotor dynamics for a single machine connected to an infinite bus can be represented as below: 
        
        %that the system response during ambient states involves many different random changes within the system.
        %Such changes include minor disturbances from system load, generation, topology and other elements.
        %They act as an excitation to the power system electromechanical dynamics, which affect the power, voltage, current, and frequency measurements produced by PMUs. 
        %For example, the rotor dynamics are implicitly coupled with the voltage and reactive power dynamics mainly introduced by exciters as (\ref{eq:withExciter}) \cite{Impedance}. 
        
            \small
                \begin{equation} \label{eq:withExciter}
                    \begin{split}               
                    & \left[\begin{array}{c}\Delta \dot{\delta} \\ \Delta \dot{\omega}\end{array}\right] 
                        = 
                        \left[\begin{array}{cc}0 & 1 \\ -\frac{V_{t} \mathrm{E}^{\prime}}{M X_{d}^{T}} \cos (\varphi) & -\frac{D}{M}\end{array}\right]\left[\begin{array}{c}\Delta \delta \\ \Delta \omega\end{array}\right] 
                        \\ & +
                        \left[\begin{array}{c c}0 & 0\\ -\frac{\mathrm{E}^{\prime}}{M X_{d}^{\prime}} \sin (\varphi) & \frac{\mathrm{V}_{t} \mathrm{E}^{\prime}}{M X_{d}^{T}} \cos (\varphi)\end{array}\right]\left[\begin{array}{c}\Delta \mathrm{V}_{t} \\ \Delta \theta_{t}\end{array}\right] 
                        \\ & + 
                        \left[\begin{array}{c}0 \\ \frac{1}{M}\end{array}\right]\left[\Delta P_{m}\right]
                        +
                        \left[\begin{array}{c}0 \\ \frac{-\mathrm{V}_{t}}{M X_{d}^{\prime}} \sin (\varphi)\end{array}\right]\left[\Delta \mathrm{E}^{\prime}\right]
                    \end{split}
                \end{equation}
            \normalsize
        where 
        $\delta$ and  $\omega$ is rotor angle and frequency;
        $M$ and $D$ are the inertia constant of the generator and  the damping coefficient of the generator; 
        $V_t$ and $\mathrm{E}^{\prime}$ denote the terminal voltage and the generator internal electromotive force (emf); 
         $X_d$ and $\varphi$ are the transient reactance and the power angle. $P_m$ is the mechanical power input through the mechanical shaft. 
        Regarding $[\Delta \mathrm{V}_{t}$ $\Delta \theta_{t}]^T$, $\Delta P_{m}$, $\Delta \mathrm{E}^{\prime}$ as input perturbations, (\ref{eq:withExciter}) can be represented as:
        \small
                \begin{equation} \label{eq:withExciter_3}
                    \begin{split}
                        \left[\begin{array}{c}\Delta \dot{\delta} \\ \Delta \dot{\omega}\end{array}\right] 
                        = & 
                        \left[\begin{array}{cc}0 & 1 \\ -\frac{V_{t} \mathrm{E}^{\prime}}{M X_{d}^{T}} \cos (\varphi) & -\frac{D}{M}\end{array}\right]\left[\begin{array}{c}\Delta \delta \\ \Delta \omega\end{array}\right] 
                        + \bm{u}
                    \end{split}
                \end{equation}
            \normalsize
        where%
            \small
                \begin{equation} \label{eq:withExciter_u}
                    \begin{split}
                        \bm{u} = & \left[\begin{array}{c c}0 & 0\\ -\frac{\mathrm{E}^{\prime}}{M X_{d}^{\prime}} \sin (\varphi) & \frac{\mathrm{V}_{t} \mathrm{E}^{\prime}}{M X_{d}^{T}} \cos (\varphi)\end{array}\right]\left[\begin{array}{c}\Delta \mathrm{V}_{t} \\ \Delta \theta_{t}\end{array}\right] 
                        + 
                        \left[\begin{array}{c}0 \\ \frac{1}{M}\end{array}\right]\left[\Delta P_{m}\right]
                        \\ +&
                        \left[\begin{array}{c}0 \\ \frac{-\mathrm{V}_{t}}{M X_{d}^{\prime}} \sin (\varphi)\end{array}\right]\left[\Delta \mathrm{E}^{\prime}\right] 
                    \end{split}
                \end{equation}
            \normalsize
        %We can reformat \ref{eq:withExciter} by introducing the effect of field flux with potential forced oscillation as external perturbation input as (\ref{eq:withExciter_3}).
        This means %the control system (excitation system and turbine governors) indeed will affect the damping ratio of natural modes as mentioned by the reviewer, while on the other hand, 
        that the forced oscillations from both excitation systems (exciters and PSSs) and turbine governors will  manifest themselves in rotor dynamics. 
        In either case, $\Delta \mathrm{E}^{\prime}$ or $\Delta P_{m}$ with forced oscillations can be represented as a Fourier series $\sum_{i=1}^{l}({x}_{i}
                        \sin \left(\omega_{F_{i}} t\right) 
                        + {y}_{i}
                        \cos \left(\omega_{F_{i}} t\right))$. 
                       Assuming the input perturbation from $[\Delta \mathrm{V}_{t}, \Delta \theta_{t}]^T$ is relatively small, (\ref{eq:withExciter_3}) can be represented approximately as:
        \small
                \begin{equation} \label{eq:withExciterSwing}
                    \begin{split}
                        \left[\begin{array}{c}\Delta \dot{\delta} \\ \Delta \dot{\omega}\end{array}\right] 
                        =
                        \left[\begin{array}{cc}0 & 1 \\ -\frac{V_{t} \mathrm{E}^{\prime}}{M X_{d}^{T}} \cos (\varphi) & -\frac{D}{M}\end{array}\right]\left[\begin{array}{c}\Delta \delta \\ \Delta \omega\end{array}\right] 
                        \\ + 
                            \sum_{i=1}^{l}\Bigl(\left[\begin{array}{c}
                        \bm{0} \\ \bm{a}_{i}
                        \end{array}\right] 
                        \sin \left(\omega_{F_{i}} t\right) 
                        + 
                        \left[\begin{array}{c}
                        \bm{0} \\ \bm{b}_{i}
                        \end{array}\right] 
                        \cos \left(\omega_{F_{i}} t\right) \Bigr)
                    \end{split}
                \end{equation}
            \normalsize  
        which is a special form of (\ref{eq:swingEquation_1})-(\ref{eq:inputFO}) for the single machine infinite bus system. %It should be noted that (\ref{eq:withExciter}) may be further extended to a multi-generation system following the procedure described in \cite{sauer2017power} Chapter 7.
        %As a result, it is believed that the associated real power oscillation (due to the turbine governor) and reactive power oscillation (due to voltage control dynamics from the excitation system) will also manifest themselves in rotor dynamics.   %may come from either exciters $\Delta \mathrm{E}^{\prime}$ or turbine governors $\Delta \tau_{m}$, while the input perturbation from $[\Delta \mathrm{V}_{t}, \Delta \theta_{t}]^T$ is relatively small.
        It should be noted that (\ref{eq:withExciter}) may be  extended to a multi-generation system following the procedure described in \cite{sauer2017power} (Chapter 7). As such, it is believed that rotor dynamics are sufficient to describe the power system with sustained forced oscillations, which is also validated in numerical studies presented in Section \ref{section:numerical study}-\ref{section:real case testing}.  %More justifications from the results perspective and the assumption used in previous works are provided in the response to the reviewer's question {<Data Driven Method> 3 on Page \color{black}12\color{black}. 
        
        \color{black}    

%%%%%%%%%%%%%%%%%%%%%%%%%%%%%%%%%%%%%%%%%%%%%%%%%%%%%%%%%%%%    
\bibliographystyle{IEEEtran}
\typeout{}
\bibliography{ref.bib}


\begin{IEEEbiography}[{\includegraphics[width=1in,height=1.25in,clip,keepaspectratio]{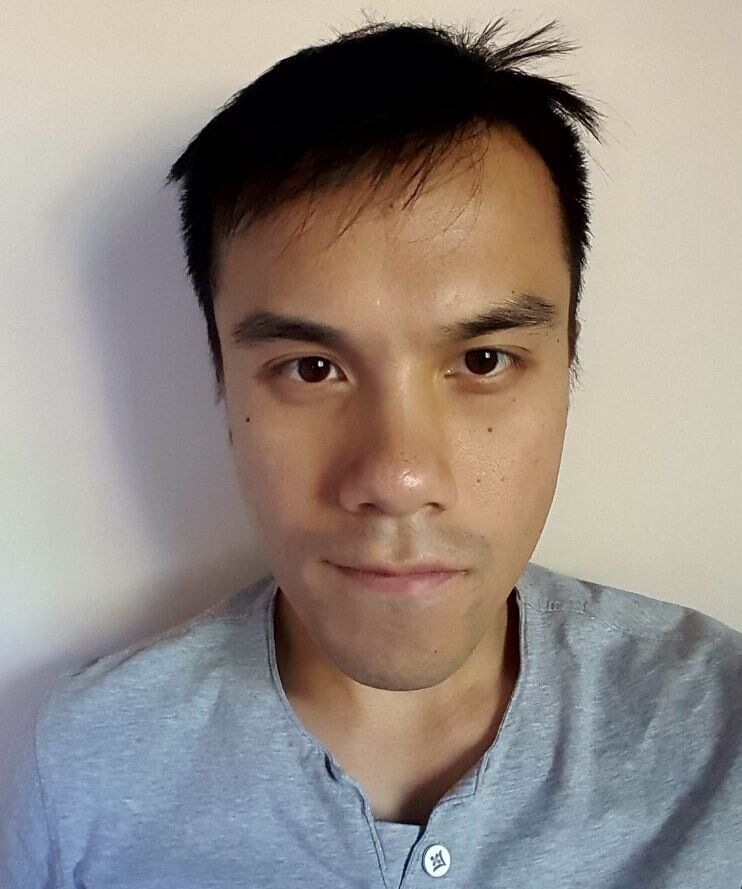}}]{Yaojie Cai}
received the the B.S. and MSc degrees in electrical and computer engineering from the University of Manitoba, Winnipeg, MB, Canada. He was a Research Assistant with Royal Military College of Canada, Kingston, ON, Canada. He is currently working toward the Ph.D. degree in electrical and computer engineering with McGill University, QC, Canada. His research interests include data analytics, power system stability and control.
\end{IEEEbiography}

\begin{IEEEbiography}[{\includegraphics[width=1in,height=1.25in,clip,keepaspectratio]{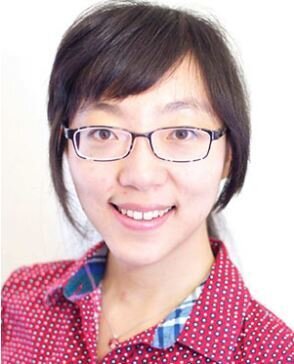}}]{Xiaozhe Wang}
(Senior Member, IEEE) is currently an Associate Professor in the Department of Electrical and Computer Engineering at McGill University, Montreal, QC, Canada. She received the Ph.D. degree in the School of Electrical and Computer Engineering from Cornell University, Ithaca, NY, USA, in 2015, and the B.S. degree in Information Science $\&$ Electronic Engineering from Zhejiang University, Zhejiang, China, in 2010. Her research interests are in the general areas of power system stability and control, uncertainty quantification in power system security and stability, and wide-area measurement system (WAMS)-based detection, estimation, and control. She is serving on the editorial boards of \textit{IEEE Transactions on Power Systems, Power Engineering Letters, and IET Generation, Transmission and Distribution}.
\end{IEEEbiography}

\begin{IEEEbiography}[{\includegraphics[width=1in,height=1.25in,clip,keepaspectratio]{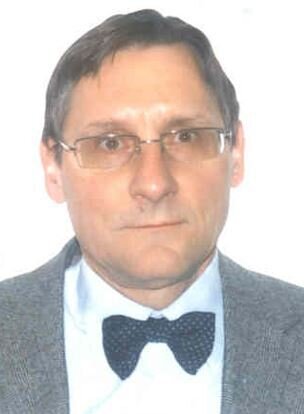}}]{Géza Joós}
(Life Fellow, IEEE) graduated from McGill University, Montreal, Canada, with an M.Eng. and Ph.D. in Electrical Engineering. He is a Professor in the Department of Electrical and Computer Engineering Department at McGill University (since 2001). He holds a Canada Research Chair in Powering Information Technologies (since 2004). His research interests are in distributed energy resources, including renewable energy resources, advanced distribution systems and microgrids. He was previously with ABB, the Université du Québec and Concordia University (Montreal, Canada). He is active in IEEE Standards Association working groups on distributed energy resources and microgrids. He is a Fellow of CIGRE, and the Canadian Academy of Engineering.
\end{IEEEbiography}

%\newpage

\begin{IEEEbiography}[{\includegraphics[width=1in,height=1.25in,clip,keepaspectratio]{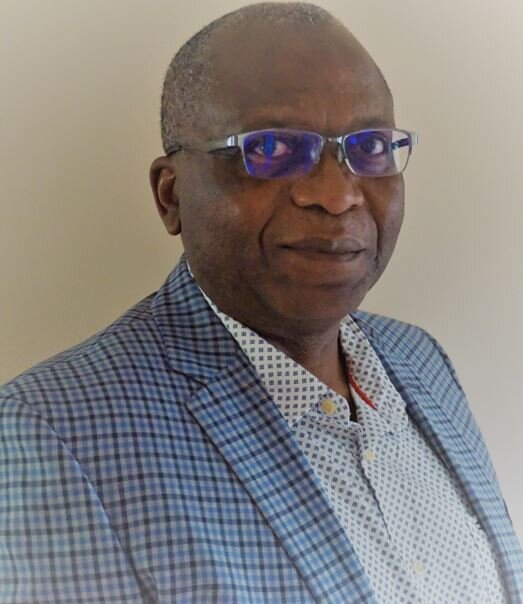}}]{Innocent Kamwa}
(Fellow, IEEE) obtained his Ph.D. in Electrical Engineering from Laval University, Quebec, QC, Canada in 1989. A full professor in the Department of Electrical Engineering and Tier 1 Canada Research Chair in Decentralized Sustainable Electricity Grids for Smart Communities at Laval University, he was previously a researcher at Hydro-Québec's Research Institute, specializing in the dynamic performance and control of power systems. He was also the Chief Scientist for Hydro-Québec's Smart Grid Innovation Program and an international consultant in power grid simulation and network stability. Dr. Kamwa is a past Editor-in-Chief of \textit{IET Generation, Transmission and Distribution}, and is currently the Editor-in-Chief of \textit{IEEE Power and Energy Magazine} and an Associate Editor of \textit{IEEE Transactions on Power Systems}. A Fellow of the Canadian Academy of Engineering and Fellow of the IEEE for his innovations in power system control, he is also the 2019 recipient of the IEEE Charles Proteus Steinmetz and Charles Concordia Awards.
\end{IEEEbiography}
\vfill

\end{document}